\documentclass[prd,onecolumn,nofootinbib,amsfonts]{revtex4-2}
\usepackage[utf8]{inputenc}
\usepackage[T1]{fontenc}
\usepackage{graphicx,bm,amsmath,color}
\usepackage[dvipsnames]{xcolor}
\usepackage{bbold}
\usepackage{wasysym}
\usepackage{verbatim}
\usepackage{amssymb}
\usepackage{hyperref}

\hypersetup{
colorlinks=true,         % false: boxed links; true: colored links, false is default
linkcolor=Orange,          % color of internal links, red is default
citecolor=OliveGreen,           % color of links to bibliography
urlcolor=Fuchsia   % color of external links, cyan is default
}

\usepackage{epstopdf}
\usepackage{animate}
\usepackage{xmpmulti}
\usepackage{centernot}
\usepackage{multirow}
\usepackage{tikz}
\usepackage{graphicx}
\usepackage{float}
\usepackage{mathtools}
\usepackage{comment}

\makeatletter

\newcommand{\be}{\begin{equation}}
\newcommand{\ee}{\end{equation}}
\newcommand{\beq}{\begin{eqnarray}}
\newcommand{\eeq}{\end{eqnarray}}
\newcommand{\ba}{\begin{align}}
\newcommand{\ea}{\end{align}}

\begin{document}

\title{Revisiting noncommutative spacetimes from the relative locality principle}
\author{J.J. Relancio}
\affiliation{\small Departamento de Matemáticas y Computación, Universidad de Burgos, Plaza Misael Bañuelos, 09001, Burgos, Spain\\Centro de Astropartículas y F\'{\i}sica de Altas Energ\'{\i}as (CAPA),
Universidad de Zaragoza, C. de Pedro Cerbuna, 12, 50009 Zaragoza, Spain}
\email{jjrelancio@ubu.es}

\begin{abstract}
Relativistic deformed kinematics leads to a loss of the absolute locality of interactions. In previous studies, some models of noncommutative spacetimes in a two-particle system that implements locality were considered. In this work, we present a characterization of the Poisson-Lie algebras formed by the noncommutative space-time coordinates of a multi-particle system and Lorentz generators as a possible restriction on these models. The relativistic deformed kinematics derived from these algebras are also discussed. Finally, we show its connection with cotangent bundle geometries.
\end{abstract}

\maketitle

\section{Introduction} A quantum gravity (QG) theory has been sought for the last decades. Several attempts have been made to combine general relativity (GR) and quantum field theory (QFT), such as loop quantum gravity~\cite{Dupuis:2012yw}, causal set theory~\cite{Wallden:2013kka}, and string theory~\cite{Mukhi:2011zz}. In the last two theories, a minimum length usually appears, changing the classical concept of spacetime in special relativity (SR).   

The introduction of a minimum length leads to nontrivial commutation rules of space-time coordinates, which can be considered a way to parametrise the quantum nature of spacetime. Noncommutative spacetimes have been considered in many down-to-top approaches to QG~\cite{Szabo:2009tn,AmelinoCamelia:2008qg}.  The concept of quantum spacetime was first proposed by Heisenberg and Ivanenko to avoid the ultraviolet divergences of the QFT. This idea was passed from Heisenberg to Peierls and Robert Oppenheimer and finally to Snyder, who published the first concrete example (Snyder model) in 1947~\cite{Snyder:1946qz}. In another vein, a curved momentum space was simultaneously considered by Born~\cite{Born:1938} to, again, avoid ultraviolet divergences in QFT. Renormalisation theories later led to the abandonment of these two concepts, but a renewed interest in QG revived these ideas in recent decades.  

In many QG approaches, relativistic deformed kinematics (RDK) have been considered as a way to capture the remnant effects of the quantum nature of spacetime (see~\cite{Addazi:2021xuf,AlvesBatista:2023wqm} for reviews). In some of these kinematics, a noncommutative spacetime has been associated with these deformations~\cite{Majid:1995qg}. Moreover, in~\cite{Carmona:2019fwf} a direct connection between a curved momentum space and RDK was shown. Furthermore, a clear relationship between the noncommutativity of space-time coordinates and a curved momentum space has been explored in~\cite{Carmona:2019fwf,Carmona:2019oph,Relancio:2021ahm,Wagner:2021thc,Wagner:2021bqz,Wagner:2022rjg,Relancio:2024nud}.

Among all the possible RDK, $\kappa$-Poincaré is one of the most studied ~\cite{Maslanka_1993,Majid1994,Zakrzewski_1994}. This Hopf algebraic deformation of the relativistic Poincaré algebra has been considered in the context of QG~\cite{AmelinoCamelia:2008qg,Addazi:2021xuf,AlvesBatista:2023wqm}. This kinematics leads to a noncommutative spacetime known as $\kappa$-Minkowski~\cite{Majid1994}. This deformation is carried out using a fixed vector, which can be timelike~\cite{Lukierski:1991pn}, lightlike ~\cite{Ballesteros:1993zi,Arratia_1998,Juric:2015hda}, or spacelike~\cite{Ballesteros_1995}. Usually, there is a deformation of the Casimir of the group (which is connected to the dispersion relation of particles, describing their free propagation), a coproduct of momenta (which we refer to as the deformed composition law of momenta, depicting the total momentum of a multi-particle system), and a coproduct of Lorentz generators (representing the Lorentz transformations of a multi-particle system).  

In~\cite{AmelinoCamelia:2011bm,AmelinoCamelia:2011pe}, a nonlocality of interactions appears when a deformed composition law of momenta is considered; only observers situated where the interaction takes place see the interaction as local. The locality of interactions can be recovered by considering a noncommutative spacetime ~\cite{Carmona:2017cry,Carmona:2019vsh,Relancio:2021ahm}. In these studies, different possibilities for implementing locality were considered, and the possible kinematics derived from them were studied (see also~\cite{Kowalski-Glikman:2014wba} for a different but complementary perspective based on rigid translations). Specifically, $\kappa$-Poincaré can be recovered in these scenarios in a simple way.

In~\cite{majid1995algebrashopfalgebrasbraided,Daszkiewicz:2007az,Lizzi:2021rlb}, the commutator of two-particle space-time coordinates is considered, which is usually performed in the context of braided tensor algebras. In this work, we start by considering the Poisson bracket of eight space-time coordinates and six Lorentz generators to study the families of Poisson-Lie algebras in 14 dimensions. These algebras are derived by enforcing Jacobi identities on an algebra that is linearly dependent on the space-time coordinates and Lorentz generators, and a fixed vector is introduced. Interestingly, for the timelike and spacelike deformations, the noncommutativity between two space-time coordinates is a mix of Snyder and $\kappa$-Minkowski, the so-called hybrid models~\cite{Meljanac:2009ej}. However, for the lightlike case, it is possible to have $\kappa$-Minkowski noncommutativity. When imposing the locality of interactions, there are several restrictions on the possible kinematics allowed depending on the model considered (\cite{Carmona:2017cry}, \cite{Carmona:2019vsh}, or \cite{Relancio:2021ahm}). Furthermore, the implementation of locality in \cite{Carmona:2017cry} can only be constructed with a lightlike deformation. We will discuss these possibilities and  their corresponding kinematics. 

Finally, we  provide a possible geometrical interpretation of the presented implementation of locality with the restrictions given by the 14 dimensional Poisson-Lie algebra. In~\cite{Relancio:2021ahm}, the phase space of a multi-particle system was considered to restrict the models that implement the locality of interactions. From the perspective of the present study, a different geometrical interpretation is provided. It is important to note that we use Poisson brackets instead of commutators because we give a geometrical interpretation at the end of the work, which is compatible with this Poisson structure. However, if we restrict ourselves to describe the 14 dimensional algebras, we can use Poisson brackets instead of commutators, as in~\cite{majid1995algebrashopfalgebrasbraided,Daszkiewicz:2007az,Lizzi:2021rlb}.   

The structure of this paper is organised as follows. In Sec.~\ref{sec:intro}, the basic concepts used in the following, such as relativistic deformed kinematics and its connection with Hopf algebras, the principle of relative locality, and the definition of line element in cotangent bundle geometries, are presented.  The construction of a 14 dimensional Poisson-Lie algebra formed by the space-time coordinates of two particles and the Lorentz generators is studied in Sec.~\ref{sec:algebra}. Subsequently, the different implementations of locality of~\cite{Carmona:2017cry}, \cite{Carmona:2019vsh}, and \cite{Relancio:2021ahm}, together with the 14 dimensional algebra conditions, are studied in Secs. \ref{sec:first}, \ref{sec:second}, and \ref{sec:general}, respectively. Then, a new geometrical interpretation of the noncommutativity of space-time coordinates is considered in Sec.~\ref{sec:geometry}. Finally, we present the conclusions and future research prospects in Sec.~\ref{sec:conclusions}. 

\section{Preliminaries on relativistic deformed kinematics and relative locality} \label{sec:intro} 
In this section, we discuss the mathematical components of relativistic deformed kinematics and their connections to the notion of relative locality and cotangent bundle geometries.

\subsection{Relativistic deformed kinematics}
Relativistic deformed kinematics are described by a deformed sum of momenta, a deformed dispersion relation, and some deformed Lorentz transformations in the two-particle system, which allow to maintain a relativity principle. Among the different models of deformed kinematics, two were primarily studied: $\kappa$-Poincaré and Snyder kinematics. The former is characterised by the introduction of a fixed vector (which can be timelike~\cite{Lukierski:1991pn}, lightlike ~\cite{Ballesteros:1993zi}, or spacelike~\cite{Ballesteros_1995}), whereas the latter preserves linear Lorentz invariance~\cite{Battisti:2010sr}. In this work, we focus on $\kappa$-Poincaré kinematics.  

These ingredients can be described using Hopf algebra nomenclature~\cite{Majid:1995qg}. The total momentum of a system of particles can be obtained from the coproduct of momenta, the deformed Lorentz transformations from the coproduct of the Lorentz generators, and the dispersion relation is modelled by the Casimir of the deformed algebra (see~\cite{Carmona:2016obd,Carmona:2017cry} for a deeper explanation). Now, we collect a simple example~\cite{Carmona:2019vsh} that we will use in the following to illustrate the relationship between coproducts, total momentum, and Lorentz transformations. The composition law reads
\be
\left(p\oplus q\right)_\mu = p_{\mu} + \left(1 + p_0/\Lambda\right) \,q_{\mu}\,,
\label{eq:dcl1}
\ee 
where $\Lambda$ is a high-energy scale (equivalently, the deformation parameter of the kinematics). The Lorentz transformations for the left particle, which are the same of those corresponding to the one-particle Lorentz transformations, are
\begin{align}
  & {\cal J}^{ij}_{L\, 0}(p) = 0\,, \quad\quad\quad {\cal J}^{ij}_{L\,k}(p) = \delta^j_k \, p_i - \delta^i_k \, p_j\,, \nonumber \\
  & {\cal J}^{0j}_{L\,0}(p) = - p_j (1+ p_0/\Lambda)\,, \quad\quad\quad {\cal J}^{0j}_{L\,k} = -\delta^j_k \left(p_0 + \left(p_0^2-\vec{p}^2\right)/2\Lambda\right) - p_j p_k/\Lambda \,,
\label{LT1}
\end{align}  
while for the right one
\begin{align}
 {\cal J}_{R\,0}^{\,0i}(p, q)\,=&\left(1+ p_0/\Lambda\right) {\cal J}_{L\,0}^{0i}(q)\,, \nonumber \\
 {\cal J}_{j}^{\,0i}(p, q)\,=&-\left(1+ p_0/\Lambda\right) {\cal J}_{L\,j}^{0i}(q)+ \,\left( \delta^i_j \vec{p}\cdot\vec{q}-p_j q_i \right)/\Lambda\,, \nonumber\\
 {\cal J}_{R\,0}^{\,ij}(p, q)\,=&\, 0\,,  \quad\quad\quad
 {\cal J}_{R\,k}^{\,ij}(p, q)=  {\cal J}_{L\,k}^{\,ij}(q)\,.
\label{calJ(2)}
\end{align}
The relativity principle is satisfied, since the total momentum of the system for two observers fulfils
\begin{equation}
\left(p\oplus q\right)^\prime_\mu=\left(p^\prime\oplus \tilde{q}\right)_\mu\,,
\label{eq:rel_princ_comp}
\end{equation}
where
\be
p^\prime_\mu=p_\mu+\epsilon_{\alpha\beta}{\cal J}^{\alpha\beta}_{L\,\mu} (p)\,,\qquad \tilde{q}_\mu=q_\mu+\epsilon_{\alpha\beta} {\cal J}^{\alpha\beta}_{R\,\mu}(p,q)  \,.
\label{eq:tilde}
\ee
The last equation can be obtained from the total Lorentz generator $J^{\mu\nu}$, defined as 
\begin{equation}
	J^{\mu\nu}  \coloneq y^\lambda {\cal J}^{\alpha\beta}_{L\,\lambda}(p)+z^\lambda  {\cal J}^{\alpha\beta}_{R\,\lambda} (p,q)\,.
\end{equation}

Then, 
\begin{equation}
	p^\prime_\mu=p_\mu+ \epsilon_{\alpha\beta}\lbrace{p_\mu,J^{\alpha\beta}\rbrace}  = p_\mu+\epsilon_{\alpha\beta}{\cal J}^{\alpha\beta}_{L\,\mu} (p)\,,
\end{equation}
and, similarly,
\begin{equation}
	q^\prime_\mu=q_\mu+ \epsilon_{\alpha\beta}\lbrace{q_\mu,J^{\alpha\beta} \rbrace} =q_\mu+\epsilon_{\alpha\beta}{\cal J}^{\alpha\beta}_{R\,\mu} (p,q)\,.
\end{equation}
Here, we introduced the Poisson brackets
\begin{equation}
	\lbrace{a,b\rbrace}=\frac{\partial a}{\partial p_\mu}\frac{\partial b}{\partial y^\mu}-\frac{\partial b}{\partial p_\mu}\frac{\partial a}{\partial y^\mu}+\frac{\partial a}{\partial q_\mu}\frac{\partial b}{\partial z^\mu}-\frac{\partial b}{\partial q_\mu}\frac{\partial a}{\partial z^\mu}\,,
\end{equation}
where $(p,y)$ and $(q,z)$ are canonically conjugated variables.

From Eq.~\eqref{eq:rel_princ_comp}, given the Lorentz transformations of Eq.~\eqref{eq:tilde}, the following equation holds~\cite{Carmona:2019fwf} 
\be
{\cal J}^{\alpha\beta}_\mu(p\oplus q) =\frac{\partial(p\oplus q)_\mu}{\partial p_\nu} {\cal J}^{\alpha\beta}_{L\,\nu} + \frac{\partial(p\oplus q)_\mu}{\partial q_\nu} {\cal J}^{\alpha\beta}_{R\,\nu}\,.
\label{eq_composition_Lorentz2}
\ee
It relates the composition law and Lorentz transformations; therefore, a relativity principle is present. 

The dispersion relation can be obtained from the Casimir of the group (invariant under Lorentz transformations), which is~\cite{Carmona:2019fwf}
\be
C(k)=\,\frac{k_0^2-\vec k^2}{1+k_0/\Lambda}\,.
\ee
Note that
\begin{equation}
    \lbrace{C(p), J^{\mu\nu}\rbrace} =\frac{\partial C(p)}{\partial p_\rho} {\cal J}^{\mu\nu}_{L\,\rho}=0\,,\qquad
    \lbrace{C(q), J^{\mu\nu}\rbrace} =\frac{\partial C(q)}{\partial q_\rho} {\cal J}^{\mu\nu}_{R\,\rho}=0\,.
    \label{eq:cplusjs}
\end{equation}

As it was shown in~\cite{Carmona:2019vsh}, this basis of the $\kappa$-Poincaré kinematics can be obtained from the well-known  bicrossproduct basis~\cite{Majid1994} considering the change of momentum basis $k_\mu \to \hat{k}_\mu$ 
\be
k_i = \hat{k}_i\,, \quad\quad\quad (1 + k_0/\Lambda) = e^{- \hat{k}_0/\Lambda}\,,
\ee
where the hatted variables correspond to those of the bicrossproduct basis of $\kappa$-Poincaré. Notice that the composition law of momenta is associative,
\be
\left(k\oplus \left(p\oplus q\right)\right)_\mu = \left(\left(k\oplus p\right)\oplus q\right)_\mu\,,
\ee 
and it is possible to extend the Lorentz generators so the relativity principle is maintained
\be
{\cal J}^{\alpha\beta}_\mu\left(k\oplus\left(p\oplus q\right)\right) =\frac{\partial\left(k\oplus\left(p\oplus q\right)\right)_\mu}{\partial k_\nu} {\cal J}^{\alpha\beta}_{L\,\nu}(k) +\frac{\partial\left(k\oplus\left(p\oplus q\right)\right)_\mu}{\partial p_\nu} {\cal J}^{\alpha\beta}_{R\,\nu}(k,p) + \frac{\partial \left(k\oplus\left(p\oplus q\right)\right)_\mu}{\partial q_\nu} {\cal J}^{\alpha\beta}_{R\,\nu}(k\oplus p,q)\,.
\label{eq_composition_Lorentz3}
\ee
This last expression can be generalised for any number of particles.

Now, we can translate these results into Hopf algebra nomenclature. The total momentum of a system of two particles will be given by the coproduct of the momenta, which, for the particular case of~\eqref{eq:dcl1}, reads
\begin{equation}
	\Delta P_\mu = P_\mu \otimes \mathrm{I}+ \left(1+ \frac{P_0}{\Lambda}\right) \otimes P_\mu\,.
\end{equation}
The coproduct of the Lorentz generators corresponding to Eqs.~\eqref{LT1} and \eqref{calJ(2)} are
\begin{equation}
	\Delta N_i = N_i \otimes \mathrm{I}+ \left(1+ \frac{P_0}{\Lambda}\right) \otimes N_i +\frac{1}{\Lambda} \epsilon_{ijk} M_j \otimes P_k\,,\qquad \Delta M_i= M_i \otimes \mathrm{I}+\mathrm{I}\otimes  M_i \,,
\end{equation}
where 
\begin{equation}
	J^{\mu\nu}=x^\rho J^{\mu\nu}_\rho\,,\qquad N_i=J^{0i}\,,\qquad M_k=\epsilon_{ijk}J^{ij}\,.
\end{equation}
Then, the relativity principle condition~\eqref{eq_composition_Lorentz2} is automatically given by the following property of Hopf algebras~\cite{Majid1994}
\begin{equation}
	[\Delta J^{\mu\nu}, \Delta P_\rho]=\Delta [J^{\mu\nu},P_\rho]\,.
\end{equation}
Moreover, 
\begin{equation}
	[\Delta J^{\mu\nu}, \Delta C]=\Delta [J^{\mu\nu},C]=0\,,\qquad \text{with}\qquad \Delta C= C \otimes \mathrm{I}+\mathrm{I} \otimes C\,.
\end{equation}
In the following, we use the notation discussed at the beginning of the section for the sake of simplicity of computation, but it is important to note that it can be translated easily into the Hopf algebra nomenclature. 

\subsection{Relative locality}
We revisit the original proposal for relative locality~\cite{AmelinoCamelia:2011bm}. We consider the action of particles which propagate freely from past infinity, interact between them, and propagate freely to future infinity, as follows:  
\begin{align}  
S^{(2)} \,=&\, \int_{-\infty}^0 d\tau \sum_{i=1,2} \left[x_{-(i)}^\mu(\tau) \dot{p}_\mu^{-(i)}(\tau) + N_{-(i)}(\tau) \left[C(p^{-(i)}(\tau)) - m_{-(i)}^2\right]\right] \nonumber \\   
& + \int^{\infty}_0 d\tau \sum_{j=1,2} \left[x_{+(j)}^\mu(\tau) \dot{p}_\mu^{+(j)}(\tau) + N_{+(j)}(\tau) \left[C(p^{+(j)}(\tau)) - m_{+(j)}^2\right]\right] \nonumber \\  
& + \xi^\mu (0)\left[{\cal P}^+_\mu(0) - {\cal P}^-_\mu(0)\right]\,,
\label{S2}  
\end{align}  
where $\dot{a}=(da/d\tau)$ denotes the derivative of variable $a$ with respect to parameter $\tau$ along the trajectory of the particle. The terms $x_{\pm(i)}$ represent the space-time coordinates of the in- and out-state particles (respectively), $p^{\pm(i)}$ are their four-momenta, and $m_{\pm(i)}$ are their masses. Furthermore, ${\cal P}^-$ (${\cal P}^+$) denotes the total four-momentum of the in-state (out-state) particles that incorporates the deformed composition law. $C(k)$ is a function that defines the deformed dispersion relation for a four-momentum $k$, and $\xi^\mu(0)$ and $N_{\pm(i)}$  are Lagrange multipliers that enforce energy-momentum conservation at the interaction as well as the dispersion relation for in- and out-state particles, respectively.  

By applying the variational principle to action \eqref{S2}, the ending (or starting) space-time coordinates of the trajectories for the in-state (out-state) particles are obtained as follows:  
\be  
x_{-(i)}^\mu(0) = \xi^\nu(0) \frac{\partial {\cal P}^-_\nu}{\partial p^{-(i)}_\mu}(0)\,, \quad\quad\quad  
x_{+(j)}^\mu(0) = \xi^\nu(0) \frac{\partial {\cal P}^+_\nu}{\partial p^{+(j)}_\mu}(0)\,.  
\label{eq_rel_loc}
\ee  
From the above equations, it is evident that only an observer located at the interaction point ($\xi^\mu(0) = 0$) will view the interaction as local, with all $x_{\pm(i)}$ coincident at zero. Although it is possible to select the Lagrange multiplier $\xi^\mu(0)$ such that the interaction appears local for one observer, other observers see it as nonlocal. This demonstrates the loss of absolute locality, an effect known as relative locality.

As discussed in~\cite{Carmona:2017cry,Carmona:2019vsh,Relancio:2021ahm}, the locality of interactions can be recovered by introducing noncommutative space–time coordinates. The general way to consider this is~\cite{Relancio:2021ahm} by means of
\be
\tilde{y}_L^\alpha = y^\mu \,\varphi^{(L)\alpha}_{(L)\mu}(p, q) + z^\mu \,\varphi^{(L)\alpha}_{(R)\mu}(p, q) {\hskip 2cm} 
\tilde{z}_R^\alpha = y^\mu \, \varphi^{(R)\alpha}_{(L)\mu}(p, q)+z^\mu\, \varphi^{(R)\alpha}_{(R)\mu}(p, q) \,.
\ee  
Since we want to recover the non-commutative spacetime for a single particle when there is only one momentum, we impose 
\begin{equation}
\varphi^{(L)\alpha}_{(L)\nu}(p,0)=\varphi^{\alpha}_{\mu}\left(p\right)\,,\qquad   \varphi^{(L)\alpha}_{(R)\nu} (0,q)= \varphi^{(R)\alpha}_{(L)\nu}(p,0)=0\,,\qquad   \varphi^{(R)\alpha}_{(R)\nu} (0,q)=\varphi^{\alpha}_{\mu}\left(q\right) \,,
\label{eq:varphi_L2}
\end{equation}
where the $\varphi$ functions lead to the noncommutativity of one particle,
\begin{equation}
	\tilde x^\mu=x^\lambda \varphi^{\mu}_{\lambda}\left(k\right)\,.
\end{equation}

The condition to have an event defined by the interaction is in this case
\be
\boxed{\varphi^\mu_\nu(p\oplus q) =\frac{\partial \left(p\oplus q\right)_\mu}{\partial p_\nu}\varphi^{(L)\alpha}_{(L)\nu}(p,q) +\frac{\partial \left(p\oplus q\right)_\mu}{\partial q_\nu} \varphi^{(L)\alpha}_{(R)\nu} (p,q) =  \frac{\partial \left(p\oplus q\right)_\mu}{\partial p_\nu}\varphi^{(R)\alpha}_{(L)\nu}(p,q) +\frac{\partial \left(p\oplus q\right)_\mu}{\partial q_\nu} \varphi^{(R)\alpha}_{(R)\nu} (p,q) } \,.
\label{loc3}
\ee

Simpler cases can be considered by restricting dependency on $\varphi$ functions. One possibility is to impose that the new space-time coordinates
of each particle are linear combinations of the space-time
coordinates of both particles, but the coefficients of the
space-time coordinates of each particle depend only on its
momentum~\cite{Carmona:2019vsh}. This implies
\begin{equation}
	\varphi^{(L)\alpha}_{(L)\nu}(p,q)=	\varphi^{\alpha}_{\nu}(p)\,,\qquad \varphi^{(L)\alpha}_{(R)\nu}(p,q)=	\varphi^{(L)\alpha}_{(R)\nu}(q)\,,\qquad 	\varphi^{(R)\alpha}_{(L)\nu}(p,q)=	\varphi^{(R)\alpha}_{(L)\nu}(p)\,,\qquad \varphi^{(R)\alpha}_{(R)\nu}(p,q)=	\varphi^{\alpha}_{\nu}(q)\,.
    \label{loc2}
\end{equation}

A different option is to impose that the new space-time coordinates do not depend on other commutative space-time coordinates ~\cite{Carmona:2017cry}. Then,
\begin{equation}
	\varphi^{(L)\alpha}_{(R)\nu}(p,q)=	0\,,\qquad 	\varphi^{(R)\alpha}_{(L)\nu}(p,q)=	0\,.
    \label{loc1}
\end{equation}

In the following sections, we consider these different possibilities and their application to our problem, which is to impose a Poisson-Lie algebra formed by the space-time coordinates of several particles and Lorentz generators.

\subsection{Geometrical interpretation}
As shown in~\cite{Carmona:2019fwf}, deformed relativistic kinematics can be obtained from a maximally symmetric momentum space: (a function of) the squared distance from the origin to a point $(k)$ leads to a dispersion relation, and the translational and Lorentz isometries of the metric give the deformed composition law and Lorentz transformations. This momentum geometry can be embedded naturally in a cotangent bundle geometry~\cite{2012arXiv1203.4101M}.

Following the idea of~\cite{2012arXiv1203.4101M}, in~\cite{Relancio:2021ahm}, a phase-space metric of a two-particle system (for flat spacetime, see~\cite{Mercati:2023pal} for its extension to curved spacetimes) was defined  
\begin{equation}
\mathcal{G}_2\,=\, G_{AB}(P) dX^A dX^B+G^{AB}(P) d P_A d P_B\,,
\label{eq:line_element_ps2}
\end{equation}
where $ G_{AB}(P) $ is an 8-dimensional metric
\begin{equation}
G_{AB}(P)\,=\,
\begin{pmatrix}
g^{LL}_{\mu\nu}(p,q) & g^{LR}_{\mu\nu}(p,q) \\
g^{RL}_{\mu\nu}(p,q) & g^{RR}_{\mu\nu}(p,q) 
\end{pmatrix}\,,
\label{eq:8-metric}
\end{equation}  
where $X^A=(y^\mu,z^\mu)$, $P_A=(p_\mu,q_\mu)$, and $A$, $B$ run from $0$ to $7$. Explicitly, this line element can be written as
\begin{equation}
\begin{split}
\mathcal{G}_2\,=\,&g^{LL}_{\mu\nu}(p,q)dy^\mu dy^\nu +2 g^{LR}_{\mu\nu}(p,q)dy^\mu dz^\nu +g^{RR}_{\mu\nu}(p,q)dz^\mu dz^\nu+\\
&g_{LL}^{\mu\nu}(p,q)dp_\mu dp_\nu +2 g_{LR}^{\mu\nu}(p,q)dp_\mu dq_\nu +g_{RR}^{\mu\nu}(p,q)dq_\mu dq_\nu\,.
\end{split}
\end{equation}
This metric can be written using an 8-dimensional tetrad~
\begin{equation}\Phi^A_B(p,q)\,=\,
\begin{pmatrix}
\varphi^{(L)\alpha}_{(L)\mu}(p,q) & \varphi^{(L)\alpha}_{(R)\mu} (p,q) \\
\varphi^{(R)\alpha}_{(L)\mu}(p,q) & \varphi^{(R)\alpha}_{(R)\mu}  (p,q)
\end{pmatrix}\,,
\label{eq:8-tetrad}
\end{equation}
such that 
	\begin{equation}
 G_{AB}(P) \,=\, \Phi^C_A(p,q)
\eta_{CD} \Phi^D_B(p,q)  \,,
\end{equation}
where 
	\begin{equation}
\eta_{CD} \,=\, \begin{pmatrix}
\eta_{\alpha\beta}&0\\
0& \eta_{\alpha\beta}
\end{pmatrix}  \,,\qquad \eta_{\alpha\beta}=\text{diag(1,-1,-1,-1)}\,,
\end{equation}
and so 
\begin{equation}
\begin{split}
g^{LL}_{\mu\nu}(p,q)\,&=\,\varphi^{(L)\alpha}_{(L)\mu}(p,q)\,\eta_{\alpha\beta}\,\varphi^{(L)\beta}_{(L)\nu}(p,q)+\varphi^{(R)\alpha}_{(L)\mu}(p,q)\,\eta_{\alpha\beta}\,\varphi^{(R)\beta}_{(L)\nu}(p,q)\,,\\
g^{LR}_{\mu\nu}(p,q)\,&=\,g^{RL}_{\nu\mu}(p,q)\,=\,\varphi^{(L)\alpha}_{(L)\mu}(p,q)\,\eta_{\alpha\beta}\,\varphi^{(L)\beta}_{(R)\nu}(p,q)+\varphi^{(R)\alpha}_{(L)\mu}(p,q)\,\eta_{\alpha\beta}\,\varphi^{(R)\beta}_{(R)\nu}(p,q)\,,\\
g^{RR}_{\mu\nu}(p,q)\,&=\,\varphi^{(L)\alpha}_{(R)\mu}(p,q)\,\eta_{\alpha\beta}\,\varphi^{(L)\beta}_{(R)\nu}(p,q)+\varphi^{(R)\alpha}_{(R)\mu}(p,q)\,\eta_{\alpha\beta}\,\varphi^{(R)\beta}_{(R)\nu}(p,q)\,.
\end{split}
\label{eq:metric_tetrad}
\end{equation}
When the composition law and Lorentz transformations in the two-particle system are required to be isometries of the metric, the momentum-dependent metric is univocally determined.

To preserve the line element and establish a relationship between incoming and outgoing coordinates/momenta through an isometry, we may introduce an intermediate state that satisfies
	\begin{equation}
 G_{AB}(P) dX^A dX^B\,=\,2 g_{\mu\nu}\left(p\oplus q\right) d\xi^\mu d\xi^\nu\,.
 \label{eq:line_element_vertex2}
\end{equation}
Multiplication by two is necessary, because we require both particles to share the same interaction vertex. Without this, the SR limit cannot be achieved when the interactions are local~\cite{Relancio:2021ahm}.
The desired relative locality conditions~\eqref{eq_rel_loc} can then be obtained using Eq.~\eqref{eq:line_element_vertex2}:
\begin{equation}
\varphi^{\alpha}_{\nu}\left(p\oplus q\right)\,=\,\frac{\partial y^\mu }{\partial \xi^\nu}\varphi^{(L)\alpha}_{(L)\mu}(p,q) +\frac{\partial z^\mu }{\partial \xi^\nu} \varphi^{(L)\alpha}_{(R)\mu} (p,q)\,=\,\frac{\partial y^\mu }{\partial \xi^\nu}\varphi^{(R)\alpha}_{(L)\mu}(p,q) +\frac{\partial z^\mu }{\partial \xi^\nu} \varphi^{(R)\alpha}_{(R)\mu} (p,q) \,,
\label{eq:tetrad_tetrads}
\end{equation}
so
\begin{equation}
\frac{\partial y^\mu }{\partial \xi^\nu}\,=\,\frac{\partial \left(p\oplus q\right)_\nu}{\partial p_\mu}\,,\qquad\frac{\partial z^\mu }{\partial \xi^\nu}\,=\,\frac{\partial \left(p\oplus q\right)_\nu}{\partial q_\mu} \,.
\label{eq:rl_flat_coord}
\end{equation}
Hence, the space-time part of line element~\eqref{eq:line_element_ps2} can be rewritten as  
\begin{equation}
ds^2_2\,=\, d\tilde{y}^\alpha \eta_{\alpha\beta} d\tilde{y}^\beta+d\tilde{z}^\alpha \eta_{\alpha\beta} d\tilde{z}^\beta \,.
\label{eq:line_element_ps2_tildes}
\end{equation}
Therefore, a connection between a noncommutative spacetime (which leads to a locality of interactions) and a geometry in the phase space is shown. In this work, we present a different connection between these two ingredients. 

\section{Closing the algebra in a multi-particle system}
\label{sec:algebra}
In this section, we start by considering the most general Poisson brackets in the two-particle system, involving the eight space-time coordinates and the six Lorentz generators, and imposing them to form a Poisson-Lie algebra. We then examine how it is possible to extend this result to any number of particles and the generic properties of the resultant Poisson brackets.

\subsection{General Poisson-Lie algebra of 14 dimensions}
We start by considering the Poisson brackets
\be \begin{split} \lbrace\tilde{y}_L^\mu, \tilde{y}_L^\nu\rbrace \,=\, \frac{c^L_L}{\Lambda}\left(\tilde{y}_L^\mu\,n^\nu-\tilde{y}_L^\nu\,n^\mu\right)+\frac{c^L_R}{\Lambda}\left(\tilde{z}_R^\mu\,n^\nu-\tilde{z}_R^\nu\,n^\mu\right) +\frac{1}{\Lambda^2}D^{\mu\nu}_{L\,\lambda\sigma}J^{\lambda  \sigma}\\ \lbrace\tilde{y}_L^\mu, \tilde{z}_R^\nu\rbrace \,=\, C_{L\,\xi}^{\mu\nu} \, \tilde{y}_L^\xi - C_{R\,\xi}^{\mu\nu} \, \tilde{z}_R^\xi +\frac{1}{\Lambda^2}D^{\mu\nu}_{\lambda\sigma}J^{\lambda  \sigma}\\ \lbrace\tilde{z}_R^\mu, \tilde{z}_R^\nu\rbrace \,=\,  \frac{c^R_L}{\Lambda}\left(\tilde{y}_L^\mu\,n^\nu-\tilde{y}_L^\nu\,n^\mu\right)+\frac{c^R_R}{\Lambda}\left(\tilde{z}_R^\mu\,n^\nu-\tilde{z}_R^\nu\,n^\mu\right) +\frac{1}{\Lambda^2}D^{\mu\nu}_{R\,\lambda\sigma}J^{\lambda  \sigma}\\ \lbrace J^{\mu\nu}, \tilde{y}_L^\nu\rbrace \,=\, \eta^{\nu\rho}\tilde{y}_L^\mu-\eta^{\mu\rho}\tilde{y}_L^\nu+\frac{1}{\Lambda}E^{\mu\nu\rho}_{L\,\lambda\sigma}J^{\lambda  \sigma}\\ \lbrace J^{\mu\nu}, \tilde{z}_R^\nu\rbrace \,=\, \eta^{\nu\rho}\tilde{z}_R^\mu-\eta^{\mu\rho}\tilde{z}_R^\nu+\frac{1}{\Lambda}E^{\mu\nu\rho}_{R\,\lambda\sigma}J^{\lambda  \sigma} \end{split} \ee where 
\be \begin{split} 
C_{L\,\rho}^{\mu\nu}=c_1^L \eta^{\mu \,\nu} n_\rho+c_2^L \delta^\mu_\rho n^\nu+c_3^L \delta^\nu_\rho n^\mu+c_4^L n^\mu n^\nu n_\rho\,,\\
C_{R\,\rho}^{\mu\nu}=c_1^R \eta^{\mu \,\nu} n_\rho+c_2^R \delta^\mu_\rho n^\nu+c_3^R \delta^\nu_\rho n^\mu+c_4^R n^\mu n^\nu n_\rho\,, \\
D^{\mu\nu}_{L\,\lambda\sigma}\,=\,j^L_1 \delta^\mu_\lambda \, \delta^\nu_\sigma +j^L_2 \left(n^\mu \delta^\nu_ \lambda n_\sigma-n^\nu \delta^\mu_ \lambda n_\lambda-n^\mu \delta^\nu_ \sigma n_\sigma+n^\nu \delta^\mu_ \sigma n_\lambda\right)\\ D^{\mu\nu}_{\lambda\sigma}\,=\,j_1 \delta^\mu_\lambda \, \delta^\nu_\sigma +j_2 \left(n^\mu \delta^\nu_ \lambda n_\sigma-n^\mu \delta^\nu_ \sigma n_\sigma\right)+j_3 \left(n^\nu \delta^\mu_ \lambda n_\sigma-n^\nu \delta^\mu_ \sigma n_\sigma\right)\\ D^{\mu\nu}_{R\,\lambda\sigma}\,=\,j^R_1 \delta^\mu_\lambda \, \delta^\nu_\sigma +j^R_2 \left(n^\mu \delta^\nu_ \lambda n_\sigma-n^\nu \delta^\mu_ \lambda n_\lambda-n^\mu \delta^\nu_ \sigma n_\sigma+n^\nu \delta^\mu_ \sigma n_\lambda\right)\\ E^{\mu\nu\rho}_{L\,\lambda\sigma}\,=\,j^L_3 \left(n^\mu \delta^\nu_ \lambda \delta^\rho_\sigma-n^\nu \delta^\mu_ \lambda \delta^\rho_\lambda-n^\mu \delta^\nu_ \sigma \delta^\rho_\sigma+n^\nu \delta^\mu_ \sigma \delta^\rho_\lambda\right)+j^L_4\left(\eta^{\mu\rho}\delta^\nu_ \lambda n_\sigma-\eta^{\nu\rho}\delta^\mu_ \sigma n_\lambda-\eta^{\mu\rho}\delta^\nu_ \lambda n_\sigma+\eta^{\nu\rho}\delta^\mu_ \sigma n_\lambda\right)+\\ j^L_5\left(n^\mu n^\rho n_\sigma \delta^\nu_\lambda-n^\nu n^\rho n_\sigma \delta^\mu_\lambda-n^\mu n^\rho n_\lambda \delta^\nu_\sigma+n^\nu n^\rho n_\lambda \delta^\mu_\sigma\right)\\ E^{\mu\nu\rho}_{R\,\lambda\sigma}\,=\,j^R_3 \left(n^\mu \delta^\nu_ \lambda \delta^\rho_\sigma-n^\nu \delta^\mu_ \lambda \delta^\rho_\lambda-n^\mu \delta^\nu_ \sigma \delta^\rho_\sigma+n^\nu \delta^\mu_ \sigma \delta^\rho_\lambda\right)+j^R_4\left(\eta^{\mu\rho}\delta^\nu_ \lambda n_\sigma-\eta^{\nu\rho}\delta^\mu_ \sigma n_\lambda-\eta^{\mu\rho}\delta^\nu_ \lambda n_\sigma+\eta^{\nu\rho}\delta^\mu_ \sigma n_\lambda\right)+\\ j^R_5\left(n^\mu n^\rho n_\sigma \delta^\nu_\lambda-n^\nu n^\rho n_\sigma \delta^\mu_\lambda-n^\mu n^\rho n_\lambda \delta^\nu_\sigma+n^\nu n^\rho n_\lambda \delta^\mu_\sigma\right)\,,\label{comyz} \end{split} \ee
$\delta^\mu_\nu$ denotes the Kronecker delta, and $n^\mu$ is a fixed (constant) vector. Of course, the Lorentz generators satisfy the algebra
\begin{equation}
	\lbrace J^{\mu\nu}, J^{\rho\sigma} \rbrace = \eta^{\nu\rho} J^{\mu\sigma} - \eta^{\mu\rho} J^{\nu\sigma} - \eta^{\nu\sigma} J^{\mu\rho} + \eta^{\mu\sigma} J^{\nu\rho}\,. 
\end{equation}
The terms without dependence on $\Lambda$ are chosen to satisfy the usual algebra in the SR limit. Imposing Jacobi identities, and depending on whether the vector is timelike  ($n^\mu=(1,0,0,0)$), lightlike ($n^\mu=(1,0,0,1)$), or spacelike ($n^\mu=(0,0,0,-1)$), we find different solutions. They can be written in a compact form as follows:
\be
\begin{split}
\lbrace\tilde{y}_L^\mu, \tilde{y}_L^\nu\rbrace =\frac{\lambda_1}{\Lambda}\left(\tilde{y}_L^\mu\,n^\nu-\tilde{y}_L^\nu\,n^\mu\right)- \frac{\alpha\, \lambda^2_1}{\Lambda^2}J^{\mu\nu}\,,\\
\lbrace\tilde{y}_L^\mu, \tilde{z}_R^\nu\rbrace = \frac{1}{\Lambda}\left(\lambda_1\tilde{z}_R^\mu\,n^\nu-\lambda_2\tilde{y}_L^\nu\,n^\mu+\eta^{\mu\nu}\left(-\lambda_1\tilde{z}_R^\alpha n_\alpha+\lambda_2 \tilde{y}_L^\alpha n_\alpha \right)\right)-\frac{\alpha\,\lambda_1\,\lambda_2}{\Lambda^2} J^{\mu\nu}\,,\\
\lbrace\tilde{z}_R^\mu, \tilde{z}_R^\nu\rbrace =  \frac{\lambda_2}{\Lambda}\left(\tilde{z}_R^\mu\,n^\nu-\tilde{z}_R^\nu\,n^\mu\right)- \frac{\alpha\,\lambda^2_2}{\Lambda^2}J^{\mu\nu}\,,\\
\lbrace J^{\mu\nu}, \tilde{y}_L^\rho\rbrace = \eta^{\nu\rho}\tilde{y}_L^\mu-\eta^{\mu\rho}\tilde{y}_L^\nu+\frac{\lambda_1}{\Lambda}\left(n^\mu J^{\nu\rho}-n^\nu J^{\mu\rho}\right)\,,\\
\lbrace J^{\mu\nu}, \tilde{z}_R^\rho\rbrace = \eta^{\nu\rho}\tilde{z}_R^\mu-\eta^{\mu\rho}\tilde{z}_R^\nu+\frac{\lambda_2}{\Lambda}\left(n^\mu J^{\nu\rho}-n^\nu J^{\mu\rho}\right)\,,
\end{split}
\ee
where $\alpha=n^\mu n_\mu$. Note that we obtain a bi-parametric (non semisimple) algebra, which indeed corresponds to the $R^{6,2}\rtimes o(3,1)$ algebra, as can be directly obtained from the following change of basis of the generators 
\begin{equation}
	\tilde y_L^\mu=y_L^\mu+\frac{\lambda_1}{\Lambda} \,n_\lambda J^{\mu\lambda}\,,\qquad\tilde z_R^\mu=z_R^\mu+\frac{\lambda_2}{\Lambda}  \,n_\lambda J^{\mu\lambda}\,.
    \label{eq:changev}
\end{equation}
This means that there is not any new Poisson-Lie algebra in 14 dimensions involving the usual coordinates and Lorentz generators than the ``trivial'' one. When the symmetry $\lambda_1=\lambda_2=1$ is imposed in every case, both particles satisfy the same algebra. Subsequently, a well-defined limit exists when the space is reduced to one particle (corresponding to $\tilde{y}_L^\mu\to 0$ or $\tilde{z}_R^\mu\to 0$). If $\lambda_1=0$ or $\lambda_2=0$, one of the coordinates  commutes but not the other, and interestingly, the Poisson brackets of $\tilde y^\mu$ and $\tilde z^\nu$ are different from zero.  Moreover, note that the lightlike case, which was also obtained in~\cite{Lizzi:2021rlb}, is the only scenario compatible with $\kappa$-Minkowski noncommutativity, because hybrid models with timelike and spacelike deformations appear.   

In the remainder of this paper, we restrict ourselves to the symmetric case. For it, one can write:
\be
\begin{split}
\lbrace\tilde{y}_L^\mu, \tilde{y}_L^\nu\rbrace = \frac{1}{\Lambda}\left(\tilde{y}_L^\mu\,n^\nu-\tilde{y}_L^\nu\,n^\mu\right)- \frac{\alpha}{\Lambda^2}J^{\mu\nu}\,,\\
\lbrace\tilde{y}_L^\mu, \tilde{z}_R^\nu\rbrace = \frac{1}{\Lambda}\left(\tilde{z}_R^\mu\,n^\nu-\tilde{y}_L^\nu\,n^\mu+\eta^{\mu\nu}\left(-\tilde{z}_R^\alpha n_\alpha+\tilde{y}_L^\alpha n_\alpha \right)\right)- \frac{\alpha}{\Lambda^2}J^{\mu\nu}\,,\\
\lbrace\tilde{z}_R^\mu, \tilde{z}_R^\nu\rbrace =  \frac{1}{\Lambda}\left(\tilde{z}_R^\mu\,n^\nu-\tilde{z}_R^\nu\,n^\mu\right)- \frac{\alpha}{\Lambda^2}J^{\mu\nu}\,,\\
\lbrace J^{\mu\nu}, \tilde{y}_L^\rho\rbrace = \eta^{\nu\rho}\tilde{y}_L^\mu-\eta^{\mu\rho}\tilde{y}_L^\nu+\frac{1}{\Lambda}\left(n^\mu J^{\nu\rho}-n^\nu J^{\mu\rho}\right)\,,\\
\lbrace J^{\mu\nu}, \tilde{z}_R^\rho \rbrace = \eta^{\nu\rho}\tilde{z}_R^\mu-\eta^{\mu\rho}\tilde{z}_R^\nu+\frac{1}{\Lambda}\left(n^\mu J^{\nu\rho}-n^\nu J^{\mu\rho}\right)\,,
\end{split}
\label{symmetric_algebra_alpha}
\ee
Hence, when there is only one particle (that is, $\tilde{y}_L^\mu\to 0$ and $p_\mu \to 0$, or $\tilde{z}_R^\mu\to 0$ and $q_\mu \to 0$), the Poisson brackets read
\be
\begin{split}
\lbrace\tilde{x}^\mu, \tilde{x}^\nu\rbrace = \frac{1}{\Lambda}\left(\tilde{x}^\mu\,n^\nu-\tilde{x}^\nu\,n^\mu\right)- \frac{\alpha}{\Lambda^2}J^{\mu\nu}\,,\\
\lbrace J^{\mu\nu}, \tilde{x}^\rho \rbrace = \eta^{\nu\rho}\tilde{x}^\mu-\eta^{\mu\rho}\tilde{x}^\nu+\frac{1}{\Lambda}\left(n^\mu J^{\nu\rho}-n^\nu J^{\mu\rho}\right)\,.
\end{split}
\label{1symmetric_algebra_alpha}
\ee

\subsection{Extension to any number of particles}
The aforementioned Poisson-Lie algebra can be extended to include any number of particles. As shown in the following equations, a (6+$4j$) Poisson-Lie algebra can be formed by considering $j$ space-time coordinates ($j>2$) and six Lorentz generators for all the possible deformations:
\be
\begin{split}
\lbrace\tilde{x}_j^\mu, \tilde{x}_j^\nu\rbrace = \frac{\lambda_j}{\Lambda}\left(\tilde{x}_j^\mu\,n^\nu-\tilde{x}_j^\nu\,n^\mu\right)- \frac{\alpha\,\lambda_j^2}{\Lambda^2}J^{\mu\nu}\,,\\ \text{for } k>j\,,\qquad 
\lbrace\tilde{x}_j^\mu, \tilde{x}_k^\nu\rbrace = \frac{1}{\Lambda}\left(\lambda_j\tilde{x}_k^\mu\,n^\nu-\lambda_k\tilde{x}_j^\nu\,n^\mu+\eta^{\mu\nu}\left(-\lambda_j \tilde{x}_k^\alpha n_\alpha+\lambda_k\tilde{x}_j^\alpha n_\alpha \right)\right)- \frac{\alpha\,\lambda_j\,\lambda_k}{\Lambda^2}J^{\mu\nu}\,, \\
\lbrace J^{\mu\nu}, \tilde{x}_j^\rho\rbrace = \eta^{\nu\rho}\tilde{x}_j^\mu-\eta^{\mu\rho}\tilde{x}_j^\nu+\frac{\lambda_j}{\Lambda}\left(n^\mu J^{\nu\rho}-n^\nu J^{\mu\rho}\right)\,,
\end{split}
\label{symmetric_algebra_alpha_n_nons}
\ee
In this case, we find the $R^{3j,j}\rtimes o(3,1)$ algebra,
since the Poisson-Lie algebra of Eq.~\eqref{symmetric_algebra_alpha_n_nons} can be obtained from the following change of generators
\begin{equation}
	\tilde x^\mu_j=x_j^\mu+\frac{\lambda_j}{\Lambda}n_\lambda J^{\mu\lambda}\,.
\end{equation}
Then, the same discussion carried out for two particles can be extended to any number of them.

For the symmetric case $\lambda_j=1$, Eq.~\eqref{symmetric_algebra_alpha_n_nons} reduces to
\be
\begin{split}
\lbrace\tilde{x}_j^\mu, \tilde{x}_j^\nu\rbrace = \frac{1}{\Lambda}\left(\tilde{x}_j^\mu\,n^\nu-\tilde{x}_j^\nu\,n^\mu\right)- \frac{\alpha}{\Lambda^2}J^{\mu\nu}\,,\\ \text{for } k>j\,,\qquad 
\lbrace\tilde{x}_j^\mu, \tilde{x}_k^\nu\rbrace = \frac{1}{\Lambda}\left(\tilde{x}_k^\mu\,n^\nu-\tilde{x}_j^\nu\,n^\mu+\eta^{\mu\nu}\left(-\tilde{x}_k^\alpha n_\alpha+\tilde{x}_j^\alpha n_\alpha \right)\right)- \frac{\alpha}{\Lambda^2}J^{\mu\nu}\,, \\
\lbrace J^{\mu\nu}, \tilde{x}_j^\rho\rbrace = \eta^{\nu\rho}\tilde{x}_j^\mu-\eta^{\mu\rho}\tilde{x}_j^\nu+\frac{1}{\Lambda}\left(n^\mu J^{\nu\rho}-n^\nu J^{\mu\rho}\right)\,.
\end{split}
\label{symmetric_algebra_alpha_n}
\ee
Finally, note that, due to the appearance of a noncommutativity, a prescription for the order of the coordinates must be imposed. This order is related to the position of the (canonically conjugated) momentum in the composition law.

\subsection{General properties and commutators}
Starting from the commutation relations in Eq.~\eqref{symmetric_algebra_alpha}, we obtain interesting results. We start by defining the relative coordinate as
\begin{equation}
	\tilde r^\mu=\tilde{y}_L^\mu-\tilde{z}_R^\mu\,.
\end{equation}
Then, the Poisson bracket of the Lorentz generators and the relative coordinate is given by
\begin{equation}
\lbrace J^{\mu\nu}, \tilde r^\rho \rbrace = \eta^{\nu\rho}\tilde r^\mu-\eta^{\mu\rho}\tilde r^\nu\,,
\label{eq:rj}
\end{equation}
which can be obtained directly from the last two conditions of~\eqref{symmetric_algebra_alpha}. 
Moreover, the Poisson bracket between the relative coordinates is
\begin{equation}
	\lbrace\tilde r^\mu, \tilde r^\nu\rbrace = \lbrace\tilde{y}_L^\mu, \tilde{y}_L^\nu\rbrace -\lbrace \tilde{y}_L^\mu, \tilde{z}_R^\nu\rbrace -\lbrace\tilde{z}_R^\mu, \tilde{y}_L^\nu\rbrace +\lbrace \tilde{z}_R^\mu, \tilde{z}_R^\nu\rbrace =0\,,
\end{equation}
as can be proven for the first three equations of~\eqref{symmetric_algebra_alpha}.
Interestingly, these relative coordinates satisfy the same algebra as in SR. In addition, for the generic implementation of locality given by Eq.~\eqref{loc3}, they satisfy 
\begin{equation}
	\begin{split}
		\lbrace\tilde r^\mu, \left(p\oplus q\right)_\nu \rbrace&=\lbrace y^\rho \,\varphi^{(L)\mu}_{(L)\rho}(p, q) + z^\rho \,\varphi^{(L)\mu}_{(R)\rho}(p, q)- y^\rho\, \varphi^{(R)\mu}_{(L)\rho}(p, q)-z^\mu\, \varphi^{(R)\mu}_{(R)\rho}(p, q),\left(p\oplus q\right)_\nu \rbrace\\
    &=-\frac{\partial (p \oplus q)_\sigma}{\partial p_\lambda} \left(\varphi^{(L)\mu}_{(L)\lambda}(p, q)-\varphi^{(R)\mu}_{(L)\lambda}(p, q)\right)-\frac{\partial (p \oplus q)_\sigma}{\partial q_\lambda} \left(\varphi^{(L)\mu}_{(R)\lambda}(p, q)-\varphi^{(R)\mu}_{(R)\lambda}(p, q)\right)
    =0\,.
    \label{eq:rcomp}
\end{split}
\end{equation}
In the last step, we used the locality condition~\eqref{loc3}. Then, there is an invariance in the relative coordinates under translations, implying that every observer agrees in the locality of the interaction. Again, this condition is the same as that obtained in SR. Therefore, even if there is a deformed Poisson-Lie algebra in the two-particle system and deformed Lorentz transformations and composition laws, leading to a noncommutative spacetime, the relative coordinates satisfy the usual Poisson-Lie algebra of SR. This fact can be understood from Eq.~\eqref{eq:changev}, since
\begin{equation}
	\tilde r^\mu=y^\mu-z^\mu\,.
\end{equation}

\section{First attempt of implementation of locality}
\label{sec:first}
For timelike and spacelike deformations, the presence of Lorentz generators in Poisson brackets that involve only $\tilde{y}_L$ and $\tilde{z}_R$ (the first three equations of~\eqref{symmetric_algebra_alpha}) implies that the simplest locality implementation with the condition~\eqref{loc1} cannot be used. However, it is possible to deal with a lightlike case because Lorentz generators do not appear. First, we consider this simple case.

As done in~\cite{Carmona:2017cry}, we introduce the functions $\phi_L$, $\phi_R$, defined by a composition law $p\oplus q$ through
\be
\phi_{L\,\sigma}^{\:\:\nu}(p, q) \,\frac{\partial(p\oplus q)_\nu}{\partial p_\rho} = \delta^\rho_\sigma\,, \qquad
\phi_{R\,\sigma}^{\:\:\nu}(p, q) \,\frac{\partial(p\oplus q)_\nu}{\partial q_\rho} = \delta^\rho_\sigma \,.
\label{eq:philr}
\ee
These functions determine the spacetime of a two-particle system once the spacetime of a one-particle system (i.e., $\varphi$) is fixed:
\be
\varphi^{(L)\sigma}_{(L)\mu}(p, q) = \phi_{L\,\sigma}^{\:\:\nu}(p, q) \,\, \varphi^\mu_\nu(p\oplus q)\,,\qquad
\varphi^{(R)\sigma}_{(R)\mu}(p, q) = \phi_{R\,\sigma}^{\:\:\nu}(p, q) \,\, \varphi^\mu_\nu(p\oplus q)\,.
\label{phiL-phiR-phi}
\ee
We note that $\phi_{L\,\sigma}^{\:\:\nu}(p, 0) = \phi_{R\,\sigma}^{\:\:\nu}(0, q) = \delta^\nu_\sigma$.

The Poisson brackets of the new spacetime coordinates in the one-particle system is given by
\be
\{\tilde{x}^\mu, \tilde{x}^\sigma\} = \{ x^\nu \varphi^\mu_\nu(k), x^\rho \varphi^\sigma_\rho(k)\} = x^\nu \frac{\partial\varphi^\mu_\nu(k)}{\partial k_\rho} \,\varphi^\sigma_\rho(k) \,-\, x^\rho \frac{\partial\varphi^\sigma_\rho(k)}{\partial k_\nu} \,\varphi^\mu_\nu(k) = x^\nu \,\left(\frac{\partial\varphi^\mu_\nu(k)}{\partial k_\rho} \,\varphi^\sigma_\rho(k) \,-\, \frac{\partial\varphi^\sigma_\nu(k)}{\partial k_\rho} \,\varphi^\mu_\rho(k)\right)\,,
\label{eq:commNCspt}
\ee
and the Poisson bracket involving the momentum by 
\be
\{k_\nu, \tilde{x}^\mu\} =\varphi^\mu_\nu(k)\,.
\ee

\subsection{Composition law}
The Poisson brackets of the space-time coordinates of the two-particle system can be computed. For the first particle one obtains
\be
\lbrace\tilde{y}_L^\mu, \tilde{y}_L^\nu\rbrace = \lbrace y^\rho \varphi^{(L)\mu}_{(L)\rho}, 
y^\sigma \varphi^{(L)\nu}_{(L)\sigma}\rbrace = y^\rho \left(\frac{\partial \varphi^{(L)\mu}_{(L)\rho}}{\partial p_\sigma}  \varphi^{(L)\nu}_{(L)\sigma} - \frac{\partial\varphi^{(L)\nu}_{(L)\rho}}{\partial p_\sigma}   \varphi^{(L)\mu}_{(L)\sigma} \right) \,,
\ee
where
\be
\frac{\partial\varphi^{\:\:\mu}_{L\rho}}{\partial p_\sigma}  \varphi^{\:\:\nu}_{L\sigma} = \left(\frac{\partial\phi_{L\,\rho}^{\:\:\lambda}}{\partial p_\sigma} \varphi^\mu_\lambda(p\oplus q) + 
\phi_{L\,\rho}^{\:\:\lambda} \frac{\partial\varphi^\mu_\lambda(p\oplus q)}{\partial(p\oplus q)_\alpha} \frac{\partial(p\oplus q)_\alpha}{\partial p_\sigma}\right) \phi_{L\,\sigma}^{\:\:\beta} \varphi^\nu_\beta(p\oplus q) \,.
\ee
Then, one obtains
\be
\begin{split}
\frac{\partial \varphi^{(L)\mu}_{(L)\rho}}{\partial p_\sigma}  \varphi^{(L)\nu}_{(L)\sigma} - \frac{\partial\varphi^{(L)\nu}_{(L)\rho}}{\partial p_\sigma}   \varphi^{(L)\mu}_{(L)\sigma}  & = 
\left(\frac{\partial\phi_{L\,\rho}^{\:\:\alpha}}{\partial p_\sigma} \phi_{L\,\sigma}^{\:\:\beta} - 
\frac{\partial\phi_{L\,\rho}^{\:\:\beta}}{\partial p_\sigma} \phi_{L\,\sigma}^{\:\:\alpha}\right) 
\varphi^\mu_\alpha(p\oplus q) \varphi^\nu_\beta(p\oplus q) \\ & + \phi_{L\,\rho}^{\:\:\alpha} \left(\frac{\partial\varphi^\mu_\alpha(p\oplus q)}{\partial(p\oplus q)_\beta} \varphi^\nu_\beta(p\oplus q) - \frac{\partial\varphi^\nu_\alpha(p\oplus q)}{\partial(p\oplus q)_\beta} \varphi^\mu_\beta(p\oplus q)  \right)\,.
\end{split}
\ee

By deriving the first equation of~\eqref{eq:philr} with respect to $p_\sigma$, one has
\be
\frac{\partial\phi_{L\,\rho}^{\:\:\alpha}}{\partial p_\sigma} = - \phi_{L\,\lambda}^{\:\:\alpha} \,\frac{\partial^2(p\oplus q)_\xi}{\partial p_\sigma \partial p_\lambda} \, \phi_{L\,\rho}^{\:\:\xi}\,,
\ee  
and, then,
\be
\frac{\partial\phi_{L\,\rho}^{\:\:\alpha}}{\partial p_\sigma} \phi_{L\,\sigma}^{\:\:\beta} = - \phi_{L\,\lambda}^{\:\:\alpha} \,\frac{\partial^2(p\oplus q)_\xi}{\partial p_\sigma \partial p_\lambda} \, \phi_{L\,\rho}^{\:\:\xi} \phi_{L\,\sigma}^{\:\:\beta}\,.
\ee
The symmetry under the exchange $\sigma\leftrightarrow \lambda$ in the second derivative of the composition law leads to a symmetry under the exchange $\alpha\leftrightarrow \beta$, and then it is easy to see that
\be
\frac{\partial\phi_{L\,\rho}^{\:\:\alpha}}{\partial p_\sigma} \phi_{L\,\sigma}^{\:\:\beta} = \frac{\partial\phi_{L\,\rho}^{\:\:\beta}}{\partial p_\sigma} \phi_{L\,\sigma}^{\:\:\alpha}\,.
\ee
This can be used to obtain a very compact expression for the space-time structure of the first particle in the two-particle system:
\be
\lbrace\tilde{y}_L^\mu, \tilde{y}_L^\nu\rbrace = y^\rho \phi_{L\,\rho}^{\:\:\alpha} \left(\frac{\partial\varphi^\mu_\alpha(p\oplus q)}{\partial(p\oplus q)_\beta} \varphi^\nu_\beta(p\oplus q) - \frac{\partial\varphi^\nu_\alpha(p\oplus q)}{\partial(p\oplus q)_\beta} \varphi^\mu_\beta(p\oplus q)\right).
\label{comy}
\ee

In the case of the coordinates of the second particle one will have
\be
\lbrace\tilde{z}_R^\mu, \tilde{z}_R^\nu\rbrace = z^\rho \phi_{R\,\rho}^{\:\:\alpha} \left(\frac{\partial\varphi^\mu_\alpha(p\oplus q)}{\partial(p\oplus q)_\beta} \varphi^\nu_\beta(p\oplus q) - \frac{\partial\varphi^\nu_\alpha(p\oplus q)}{\partial(p\oplus q)_\beta} \varphi^\mu_\beta(p\oplus q)\right).
\label{comz}
\ee

The remaining space-time Poisson bracket in the two-particle system is
\be
\lbrace\tilde{y}_L^\mu, \tilde{z}_R^\nu\rbrace = \lbrace y^\rho  \varphi^{(L)\mu}_{(L)\rho}, 
z^\sigma  \varphi^{(R)\nu}_{(R)\sigma}\rbrace = y^\rho \frac{\partial \varphi^{(L)\mu}_{(L)\rho}}{\partial q_\sigma}   \varphi^{(R)\nu}_{(R)\sigma} - z^\sigma \frac{\partial  \varphi^{(R)\nu}_{(R)\sigma}}{\partial p_\rho} \varphi^{(L)\mu}_{(L)\rho}\,.
\label{eq:comyz}
\ee

Moreover, if one considers a function $\varphi^\mu_\nu$ such that
\be
\left(\frac{\partial\varphi^\mu_\nu(p)}{\partial p_\rho} \,\varphi^\sigma_\rho(p) \,-\, \frac{\partial\varphi^\sigma_\nu(p)}{\partial p_\rho} \,\varphi^\mu_\rho(p)\right) = C^{\mu\sigma}_\lambda \varphi^\lambda_\nu(p)
\label{kappa}
\ee
where 
\begin{equation}
	C^{\mu\sigma}_\lambda=\delta^\mu_\lambda n^\sigma-\delta^\sigma_\lambda n^\mu\,,
\end{equation}
then the space-time algebra for one particle (see \eqref{eq:commNCspt}) reduces to
\be
\lbrace\tilde{x}^\mu, \tilde{x}^\sigma\rbrace = C^{\mu\sigma}_\lambda \tilde{x}^\lambda\,, 
\ee  
and 
\be
\lbrace\tilde{y}_L^\mu, \tilde{y}_L^\nu\rbrace = y^\rho \phi_{L\,\rho}^{\:\:\alpha} C^{\mu\nu}_\lambda \varphi^\lambda_\alpha(p\oplus q) = C^{\mu\nu}_\lambda \tilde{y}_L^\lambda\,,\qquad
\lbrace\tilde{z}_R^\mu, \tilde{z}_R^\nu\rbrace = z^\rho \phi_{R\,\rho}^{\:\:\alpha} C^{\mu\nu}_\lambda \varphi^\lambda_\alpha(p\oplus q) = C^{\mu\nu}_\lambda \tilde{z}_R^\lambda\,,
\ee
which is the same algebra as that of the one-particle system.
This simple relationship between the space-time structures of a two-particle system and that of a one-particle system is consistent with the property that a two-particle system reduces to a one-particle system when one of the momenta is zero. 

In the general case, one does not have a Poisson-Lie algebra in the two-particle spacetime owing to the Poisson brackets \eqref{eq:comyz}. Such an algebra can be added to limit the wide range of possible solutions. We first note that
\be
\begin{split}
\frac{\partial\varphi^{(L)\mu}_{(L)\rho}}{\partial q_\sigma}   \varphi^{(R)\nu}_{(R)\sigma}\,=& 
\left(\frac{\partial\phi_{L\,\rho}^{\:\:\alpha}}{\partial q_\sigma} \varphi^\mu_\alpha(p\oplus q) +  \phi_{L\,\rho}^{\:\:\alpha} \frac{\partial\varphi^\mu_\alpha(p\oplus q)}{\partial(p\oplus q)_\lambda} \frac{\partial(p\oplus q)_\lambda}{\partial q_\sigma}\right)  \phi_{R\,\sigma}^{\:\:\beta} \varphi^\nu_\beta(p\oplus q) \\ =& \phi_{L\,\rho}^{\:\:\alpha} \left[- \phi_{L\,\lambda}^{\:\:\gamma} \phi_{R\,\sigma}^{\:\:\beta} \frac{\partial^2(p\oplus q)_\alpha}{\partial q_\sigma \partial p_\lambda} \varphi^\mu_\gamma(p\oplus q) + \frac{\partial\varphi^\mu_\alpha(p\oplus q)}{\partial(p\oplus q)_\beta}\right] \varphi^\nu_\beta(p\oplus q)\,,
\end{split}
\ee
where we used the relation between the derivatives of a matrix and its inverse. Similarly (exchanging $L\leftrightarrow R$ and $\mu\leftrightarrow\nu$),
\be
\frac{\partial  \varphi^{(R)\nu}_{(R)\sigma}}{\partial p_\rho} \varphi^{(L)\mu}_{(L)\rho} = 
 \phi_{R\,\sigma}^{\:\:\alpha} \left[- \phi_{R\,\lambda}^{\:\:\gamma} \phi_{L\,\rho}^{\:\:\beta} \frac{\partial^2(p\oplus q)_\alpha}{\partial p_\rho \partial q_\lambda} \varphi^\nu_\gamma(p\oplus q) + \frac{\partial\varphi^\nu_\alpha(p\oplus q)}{\partial(p\oplus q)_\beta}\right] \varphi^\mu_\beta(p\oplus q)\,.
\ee
Then  \eqref{eq:comyz} can be expressed as 
\be
\begin{split}
\lbrace\tilde{y}_L^\mu, \tilde{z}_R^\nu\rbrace = &y^\rho \,\phi_{L\,\rho}^{\:\:\alpha} \left[- \phi_{L\,\lambda}^{\:\:\gamma} \phi_{R\,\sigma}^{\:\:\beta} \frac{\partial^2(p\oplus q)_\alpha}{\partial q_\sigma \partial p_\lambda} \varphi^\mu_\gamma(p\oplus q) + \frac{\partial\varphi^\mu_\alpha(p\oplus q)}{\partial(p\oplus q)_\beta}\right] \varphi^\nu_\beta(p\oplus q) \\ -&z^\sigma  \,\phi_{R\,\sigma}^{\:\:\alpha} \left[- \phi_{R\,\lambda}^{\:\:\gamma} \phi_{L\,\rho}^{\:\:\beta} \frac{\partial^2(p\oplus q)_\alpha}{\partial p_\rho \partial q_\lambda} \varphi^\nu_\gamma(p\oplus q) + \frac{\partial\varphi^\nu_\alpha(p\oplus q)}{\partial(p\oplus q)_\beta}\right] \varphi^\mu_\beta(p\oplus q)\,.
\end{split}
\ee

If one wants to have an eight-dimensional Poisson-Lie algebra with the space-time coordinates of the two-particle system as generators, the composition law has to be such that
\be
\begin{split}
  & \left[- \phi_{L\,\lambda}^{\:\:\gamma} \phi_{R\,\sigma}^{\:\:\beta} \frac{\partial^2(p\oplus q)_\alpha}{\partial q_\sigma \partial p_\lambda} \varphi^\mu_\gamma(p\oplus q) + \frac{\partial\varphi^\mu_\alpha(p\oplus q)}{\partial(p\oplus q)_\beta}\right] \varphi^\nu_\beta(p\oplus q) = C_{L\,\xi}^{\mu\nu} \, \varphi^\xi_\alpha(p\oplus q)\,, \\
  & \left[- \phi_{R\,\lambda}^{\:\:\gamma} \phi_{L\,\rho}^{\:\:\beta} \frac{\partial^2(p\oplus q)_\alpha}{\partial p_\rho \partial q_\lambda} \varphi^\nu_\gamma(p\oplus q) + \frac{\partial\varphi^\nu_\alpha(p\oplus q)}{\partial(p\oplus q)_\beta}\right] \varphi^\mu_\beta(p\oplus q) =
  C_{R\,\xi}^{\mu\nu} \, \varphi^\xi_\alpha(p\oplus q)\,,
\end{split}
\ee
and then
\be
\lbrace\tilde{y}_L^\mu, \tilde{z}_R^\nu\rbrace = C_{L\,\xi}^{\mu\nu} \, \tilde{y}_L^\xi - C_{R\,\xi}^{\mu\nu} \, \tilde{z}_R^\xi\,,
\ee
where
\begin{equation}
	C_{L\,\xi}^{\mu\nu} =C_{R\,\xi}^{\nu\mu} =-\delta^\nu_\xi n^\mu+\eta^{\mu\nu}n_\xi\,,
    \label{eq:clcr}
\end{equation}
are given by Eq.~\eqref{1symmetric_algebra_alpha}.

The conditions on the composition law in order to have a Poisson-Lie algebra with space-time coordinates as generators can be written as two relations for the second derivatives of the composition law in terms of the first derivatives,
\be
\begin{split}
  & \frac{\partial^2(p\oplus q)_a}{\partial p_b \partial q_c} = \bar{\varphi}^d_\mu \,\left[\frac{\partial\varphi^\mu_a}{\partial(p\oplus q)_e} - C_{L\,\xi}^{\mu\nu} \,\varphi^\xi_a\,\bar{\varphi}^e_\nu\right] \,\frac{\partial(p\oplus q)_d}{\partial p_b} \, \frac{\partial(p\oplus q)_e}{\partial q_c} \,,\\
  & \frac{\partial^2(p\oplus q)_a}{\partial p_b \partial q_c} = \bar{\varphi}^d_\mu \,\left[\frac{\partial\varphi^\mu_a}{\partial(p\oplus q)_e} - C_{R\,\xi}^{\nu\mu} \,\varphi^\xi_a\,\bar{\varphi}^e_\nu\right] \,\frac{\partial(p\oplus q)_e}{\partial p_b} \, \frac{\partial(p\oplus q)_d}{\partial q_c} \,,
  \label{eq:difeq_momentum}
\end{split}
\ee
where $\bar{\varphi}$ denotes the inverse of $\varphi$.
The compatibility of these two relations requires that
\be
\bar{\varphi}^d_\mu \,\left[\frac{\partial\varphi^\mu_a}{\partial(p\oplus q)_e} - C_{L\,\xi}^{\mu\nu} \,\varphi^\xi_a\,\bar{\varphi}^e_\nu\right] = \bar{\varphi}^e_\mu \,\left[\frac{\partial\varphi^\mu_a}{\partial(p\oplus q)_d} - C_{R\,\xi}^{\nu\mu} \,\varphi^\xi_a\,\bar{\varphi}^d_\nu\right] \,,
\label{eq:63}
\ee
or, equivalently,
\be
\bar{\varphi}^d_\mu \,\frac{\partial\varphi^\mu_a}{\partial(p\oplus q)_e} \,-\,
\bar{\varphi}^e_\mu \,\frac{\partial\varphi^\mu_a}{\partial(p\oplus q)_d} =   \varphi^\xi_a \,\left(C_{L\,\xi}^{\mu\nu}\,\bar{\varphi}^d_\mu \,\bar{\varphi}^e_\nu - C_{R\,\xi}^{\nu\mu} \,\bar{\varphi}^e_\mu \,\bar{\varphi}^d_\nu\right) = \left(C_{L\,\xi}^{\mu\nu} -  C_{R\,\xi}^{\mu\nu}\right) \,\varphi^\xi_a \,\bar{\varphi}^d_\mu \,\bar{\varphi}^e_\nu\,.
\ee
If we multiply both sides by $\varphi^\rho_d \,\varphi^\sigma_e$ we find
\be
\frac{\partial\varphi^\rho_a}{\partial(p\oplus q)_e} \,\varphi^\sigma_e \,-\, \frac{\partial\varphi^\sigma_a}{\partial(p\oplus q)_d} \,\varphi^\rho_d =  \left(C_{L\,\xi}^{\rho\sigma} -  C_{R\,\xi}^{\rho\sigma}\right) \,\varphi^\xi_a\,.
\label{eq:65}
\ee
This is a relation between the two-particle spacetime algebra, i.e., between the coefficients ($C_L$, $C_R$), and the functions $\varphi$ defining a noncommutative one-particle spacetime such that
\be
\lbrace\tilde{x}^\mu, \tilde{x}^\nu\rbrace = \left(C_{L\,\rho}^{\mu\nu} -  C_{R\,\rho}^{\mu\nu}\right) \,\tilde{x}^\rho=C_{\rho}^{\mu\nu} \,\tilde{x}^\rho\,.
\label{eq:condition_Cs}
\ee
This relation can be heuristically obtained starting from the commutator of $\tilde{y}_L$ and $\tilde{z}_R$ and identifying $\tilde{y}_L=\tilde{z}_R=\tilde{x}$.

For simplicity, we express the differential equation~\eqref{eq:difeq_momentum} for the composition law of momenta as 
\begin{equation}
\frac{\partial^2 (p \oplus q)_a}{\partial p_b \partial q_c} + \Gamma^{de}_a(p \oplus q) \frac{\partial (p \oplus q)_d}{\partial p_b} \frac{\partial (p \oplus q)_e}{\partial q_c} = 0\,,
\label{eq:composition_law}
\end{equation}
where
\begin{equation}
\Gamma^{de}_a(p \oplus q) \equiv \phi^d_\mu \frac{\partial \varphi^\mu_a}{\partial (p \oplus q)_e} + C^{\mu \nu}_{L \,\xi} \varphi^\xi_a \phi^d_\mu \phi^e_\nu\,.
\label{eq:gamma_definition}
\end{equation}
It is interesting to note that the same relationship between the composition law (involving some $\Gamma$ coefficients) was obtained from a geometrical perspective in~\cite{AmelinoCamelia:2011bm} and~\cite{Amelino-Camelia:2013sba}. 

Since we are considering the algebra of Eq.~\eqref{symmetric_algebra_alpha}, and taking into account Eq.~\eqref{eq:clcr}, from Eq.~\eqref{eq:difeq_momentum} one finds
\begin{equation}
\Gamma^{de}_a(k) =\Gamma^{ed}_a(k)\,.
\label{eq:193}
\end{equation}
Therefore, the solution of Eq.~\eqref{eq:composition_law} is a commutative composition law. Moreover, we can prove that it must be symmetric. This can be achieved by taking the limit when one of the momenta approaches zero in Eq.~\eqref{eq:composition_law}.
\begin{equation}
	\lim_{p\to0} \frac{\partial^2 (p \oplus q)_a}{\partial p_b \partial q_c} + \Gamma^{de}_a(p \oplus q) \frac{\partial (p \oplus q)_d}{\partial p_b} \frac{\partial (p \oplus q)_e}{\partial q_c} = \frac{\partial L^b_a(q)}{\partial q_c} + \Gamma^{db}_a(q)   L^b_d(q)= 0\,.
    \label{eq:limcom}
\end{equation}
Thus, the generators
\begin{equation}
	L^b_a(q)=\lim_{p\to0} \frac{\partial (p \oplus q)_a}{\partial p_b}
\end{equation}
lead to the infinitesimal left translations of the composition law, i.e., if we define
\begin{equation}
	T_L^\mu=z^\nu L^\mu_\nu(q)\,,
\end{equation}
then
\begin{equation}
	(\epsilon \oplus q)_\mu=q_\mu+\epsilon_\nu\lbrace{q_\mu,T^\nu_L\rbrace}=\epsilon_\nu L^\nu_\mu (q)+q_\mu\,.
\end{equation}
Hence, multiplying~\eqref{eq:limcom} by $L^e_c(q)$  we find
\begin{equation}
 \frac{\partial L^b_a(q)}{\partial q_c} L^e_c(q)=-  \Gamma^{cd}_a(q)   L^b_d(q) L^e_c(q)\,.
\end{equation}
This implies
\begin{equation}
 \frac{\partial L^b_a(q)}{\partial q_c} L^e_c(q)- \frac{\partial L^e_a(q)}{\partial q_c} L^b_c(q) = 0\,,
\end{equation}
where we used the symmetry~\eqref{eq:193} of the $\Gamma$ coefficients. This means that the generators of left translations satisfy
\begin{equation}
\lbrace{T^\mu_L,T^\nu_L\rbrace} = 0\,.
\end{equation}
Similarly, one can find the same property for the generators of the right translations:
\begin{equation}
\lbrace{T^\mu_R,T^\nu_R\rbrace} = 0\,,
\end{equation}
where
\begin{equation}
	T_R^\mu=y^\nu R^\mu_\nu(p)\,,\qquad R^b_a(p)=\lim_{q\to0} \frac{\partial (p \oplus q)_a}{\partial q_b}\,.
\end{equation}

Therefore, we know that the composition law must be symmetric and associative; that is, it can be obtained from a change of momentum basis from the sum (as done in~\cite{Judes:2002bw}):
\begin{equation}
	p_\mu^\prime=f_\mu (p)\qquad\implies\qquad (p^\prime \oplus q^\prime)_\mu = f^{-1}_\mu\left(f_\mu(p)+f_\mu(q)\right)\,.
\end{equation}

We must select the functions $\varphi$ that reproduce the Poisson bracket~\eqref{kappa} to obtain the composition law. Since there are infinite options, we choose the simple one considered in~\cite{Carmona:2019vsh,Relancio:2021ahm},
\begin{equation}
	\varphi^\mu_\nu=\delta^\mu_\nu \left(1+\frac{p n}{\Lambda}\right)\,,
    \label{eq:varphi}
\end{equation}
with $p n=p_\mu n^\mu$. By solving the system of second-order partial differential equations given by Eq.~\eqref{symmetric_algebra_alpha}, one finds
\begin{equation}
	(p \oplus q)_\mu=p_\mu f_1 +q_\mu f_2 + \Lambda n_\mu \left(f_3 \frac{p^2}{\Lambda^2}+f_4 \frac{pq}{\Lambda^2}+f_5 \frac{q^2}{\Lambda^2}\right)\,,
\end{equation}
where
\begin{align}
		f_1&=\frac{1+q n/\Lambda}{1-pn\, qn/\Lambda^2}\,,\qquad f_2=\frac{1+p n/\Lambda}{1-pn\, qn/\Lambda^2}\,,\qquad f_3=-\frac{q n/\Lambda}{2\left(1-pn\, qn/\Lambda^2\right)}\,,\qquad f_4=-\frac{1}{1-pn\, qn/\Lambda^2}\,,\\\qquad f_5&=-\frac{p n/\Lambda}{2\left(1-pn\, qn/\Lambda^2\right)}\,,\qquad k^2=k_\mu \eta^{\mu\nu}k_\nu\,,\qquad pq=p_\mu \eta^{\mu\nu}q_\nu\,.
\end{align}

\subsection{Lorentz transformations}
The Lorentz generators of the one-particle system can be obtained from the set of differential equations of the Poisson bracket
\begin{align}
\lbrace{J^{\mu\nu},\tilde{x}^\rho\rbrace}=&\lbrace{x^\lambda  {\cal J}^{\mu\nu}_{\lambda},x^\sigma \varphi^{\rho}_{\sigma}\rbrace}=x^\lambda\left(\frac{\partial {\cal J}^{\mu\nu}_{\lambda}}{\partial k_\sigma}\varphi^{\rho}_{\sigma}-\frac{\partial \varphi^{\rho}_{\lambda}}{\partial k_\sigma}{\cal J}^{\mu\nu}_{\sigma}\right)\,,
    \label{eq:jvarphi}
\end{align}
obtaining, for the $\varphi$ functions of~\eqref{eq:varphi}, 
\begin{equation}
	J^{\mu\nu}_\rho(k)=\left(  \delta^{\mu}_{\rho} k^{\nu} -\delta^{\nu}_{\rho} k^{\mu}\right) 
    + \left( \delta^{\nu}_{\rho} n^{\mu} - \delta^{\mu}_{\rho} n^{\nu} \right) 
    \left( \frac{k^2/\Lambda}{2 (1 + kn/\Lambda)} \right)\,.
    \label{eq:jlight}
\end{equation}

To obtain the Lorentz generators, we start by considering the following Poisson brackets
\begin{align}
\lbrace{J^{\mu\nu},\tilde{y}_L^\rho\rbrace}=&\lbrace{y^\lambda  {\cal J}^{\mu\nu}_{L\,\lambda}+z^\lambda {\cal J}^{\mu\nu}_{R\,\lambda} ,y^\sigma \varphi^{(L)\rho}_{(L)\sigma}\rbrace}=y^\lambda\left(\frac{\partial {\cal J}^{\mu\nu}_{L\,\lambda}}{\partial p_\sigma}\varphi^{(L)\rho}_{(L)\sigma}-\frac{\partial \varphi^{(L)\rho}_{(L)\lambda}}{\partial p_\sigma}{\cal J}^{\mu\nu}_{L\,\sigma} -\frac{\partial \varphi^{(L)\rho}_{(L)\lambda}}{\partial q_\sigma}{\cal J}^{\mu\nu}_{R\,\sigma}\right)+z^\lambda \frac{\partial {\cal J}^{\mu\nu}_{R\,\lambda}}{\partial p_\sigma}\varphi^{(L)\rho}_{(L)\sigma} 
\label{eq:jy1}\\
    =& y^\lambda\left(\eta^{\nu\rho}\varphi^{(L)\mu}_{(L)\lambda}-\eta^{\mu\rho}\varphi^{(L)\nu}_{(L)\lambda}+\frac{1}{\Lambda}\left(n^\mu  {\cal J}^{\nu\rho}_{L\,\lambda}-n^\nu  {\cal J}^{\mu\rho}_{L\,\lambda}\right)\right)+z^\lambda \frac{1}{\Lambda}\left(n^\mu  {\cal J}^{\nu\rho}_{R\,\lambda}-n^\nu  {\cal J}^{\mu\rho}_{R\,\lambda}\right)\,,
    \label{eq:jy2}
\end{align}
where, in the last equality, we used the fourth equation of~\eqref{symmetric_algebra_alpha}. We can now expand the first term of the last equality in Eq.~\eqref{eq:jy1}:
\begin{equation}
	\begin{split}
	    -\frac{\partial \varphi^{(L)\rho}_{(L)\lambda}}{\partial p_\sigma}{\cal J}^{\mu\nu}_{L\,\sigma}=&-{\cal J}^{\mu\nu}_{L\,\sigma}\left(\frac{\partial \phi_{L\,\lambda}^{\:\:\alpha} }{\partial p_\sigma} \varphi^\rho_\alpha(p \oplus q)+\phi_{L\,\lambda}^{\:\:\alpha} \frac{\partial\varphi^\rho_\alpha(p \oplus q)}{\partial(p \oplus q)_\tau}\frac{\partial (p \oplus q)_\tau}{\partial p_\sigma}\right)=\\
        =&{\cal J}^{\mu\nu}_{L\,\sigma}\left(\phi_{L\,\lambda}^{\:\:\xi}\frac{\partial^2 (p \oplus q)_\xi}{\partial p_\sigma \partial p_\tau  }\varphi^\rho_\alpha (p \oplus q) \phi_{L\,\tau}^{\:\:\alpha}-\phi_{L\,\lambda}^{\:\:\alpha} \frac{\partial\varphi^\rho_\alpha(p \oplus q)}{\partial(p \oplus q)_\tau}\frac{\partial (p \oplus q)_\tau}{\partial p_\sigma}\right)\,,\\
         -\frac{\partial \varphi^{(L)\rho}_{(L)\lambda}}{\partial q_\sigma}{\cal J}^{\mu\nu}_{R\,\sigma}=&-{\cal J}^{\mu\nu}_{R\,\sigma}\left(\frac{\partial \phi_{L\,\lambda}^{\:\:\alpha} }{\partial q_\sigma} \varphi^\rho_\alpha(p \oplus q)+\phi_{L\,\lambda}^{\:\:\alpha} \frac{\partial\varphi^\rho_\alpha(p \oplus q)}{\partial(p \oplus q)_\tau}\frac{\partial (p \oplus q)_\tau}{\partial q_\sigma}\right)=\\
        =&{\cal J}^{\mu\nu}_{L\,\sigma}\phi_{L\,\lambda}^{\:\:\xi}\frac{\partial^2 (p \oplus q)_\xi}{\partial p_\tau  \partial q_\sigma }\varphi^\rho_\alpha (p \oplus q) \phi_{L\,\tau}^{\:\:\alpha}-\phi_{L\,\lambda}^{\:\:\alpha} \frac{\partial\varphi^\rho_\alpha(p \oplus q)}{\partial(p \oplus q)_\tau}\frac{\partial (p \oplus q)_\tau}{\partial q_\sigma}\left(J^{\mu\nu}_\tau(p \oplus q)-\frac{\partial (p \oplus q)_\tau}{\partial p_\sigma}J^{\mu\nu}_{L\,\sigma}\right)\phi_{L\,\lambda}^{\:\:\alpha}\,.
	\end{split}
\end{equation}
where we used Eq.~\eqref{eq_composition_Lorentz2} and 
\begin{equation}
\frac{\partial\phi_{L\,\rho}^{\:\:\alpha}}{\partial q_\sigma} = - \phi_{L\,\lambda}^{\:\:\alpha} \,\frac{\partial^2(p\oplus q)_\xi}{ \partial p_\lambda \partial q_\sigma} \, \phi_{L\,\rho}^{\:\:\xi}\,.
\end{equation}
Then, we find
\begin{equation}
	\begin{split}
			&y^\lambda\left(\frac{\partial {\cal J}^{\mu\nu}_{L\,\lambda}}{\partial p_\sigma}\varphi^{(L)\rho}_{(L)\sigma}-\frac{\partial \varphi^{(L)\rho}_{(L)\lambda}}{\partial p_\sigma}{\cal J}^{\mu\nu}_{L\,\sigma} -\frac{\partial \varphi^{(L)\rho}_{(L)\lambda}}{\partial q_\sigma}{\cal J}^{\mu\nu}_{R\,\sigma}\right)\\
        =&y^\lambda\left(\frac{\partial {\cal J}^{\mu\nu}_{L\,\lambda}}{\partial p_\sigma}\varphi^{(L)\rho}_{(L)\sigma}+\phi_{L\,\lambda}^{\:\:\alpha} \varphi^{(L)\rho}_{(L)\tau}\left(\frac{\partial^2(p \oplus q)_\alpha}{\partial p_\sigma \partial p_\tau} {\cal J}^{\mu\nu}_{L\,\sigma}+\frac{\partial^2(p \oplus q)_\alpha}{\partial q_\sigma \partial p_\tau} {\cal J}^{\mu\nu}_{R\,\sigma}\right)-\frac{\partial \varphi^{\rho}_{\alpha}(p \oplus q)}{\partial (p \oplus q)_\tau}{\cal J}^{\mu\nu}_{\tau} (p \oplus q)\phi_{L\,\lambda}^{\:\:\alpha} \right)\\
        =&y^\lambda \phi_{L\,\lambda}^{\:\:\alpha} \left(\varphi^\rho_\xi (p \oplus q)\frac{\partial{\cal J}^{\mu\nu}_{\alpha} (p \oplus q)}{\partial (p \oplus q)_\xi}-\frac{\partial \varphi^\rho_\alpha (p \oplus q)}{\partial (p \oplus q)_\xi} {\cal J}^{\mu\nu}_{\xi}(p \oplus q)-\frac{\partial (p \oplus q)_\alpha}{\partial p_\xi }\frac{\partial {\cal J}^{\mu\nu}_{R\,\xi}}{\partial p_\tau}\varphi^{(L)\rho}_{(L)\tau}\right)\,.
\end{split}
\end{equation}

The first two terms of the last equation correspond to the Poisson bracket in the one-particle system between the Lorentz generators and the noncommutative space-time coordinates. Then, the previous equation can be written as
\begin{equation}
	\begin{split}
	    		&y^\lambda \phi_{L\,\lambda}^{\:\:\alpha} \left(\eta^{\nu\rho}\varphi^\mu_\alpha (p \oplus q)-\eta^{\mu\rho}\varphi^\nu_\alpha (p \oplus q)+\frac{1}{\Lambda}\left(n^\mu  {\cal J}^{\nu\rho}_{\lambda}-n^\nu  {\cal J}^{\mu\rho}_{\lambda}\right)-\frac{\partial (p \oplus q)_\alpha}{\partial p_\xi }\frac{\partial {\cal J}^{\mu\nu}_{R\,\xi}}{\partial p_\tau}\varphi^{(L)\rho}_{(L)\tau}\right)\\
    =&y^\lambda\left(\eta^{\nu\rho}\varphi^{(L)\mu}_{(L)\lambda}-\eta^{\mu\rho}\varphi^{(L)\nu}_{(L)\lambda}+\frac{1}{\Lambda}\left(n^\mu  {\cal J}^{\nu\rho}_{L\,\lambda}-n^\nu  {\cal J}^{\mu\rho}_{L\,\lambda}\right)+\frac{1}{\Lambda}\left(n^\mu  {\cal J}^{\nu\rho}_{R\,\lambda}-n^\nu  {\cal J}^{\mu\rho}_{R\,\lambda}\right)-\frac{\partial {\cal J}^{\mu\nu}_{R\,\lambda}}{\partial p_\tau}\varphi^{(L)\rho}_{(L)\tau}\right)\,.
	\end{split}
\end{equation}

Then, for satisfying Eq.~\eqref{eq:jy2}, the following equation must hold
\begin{equation}
	\frac{\partial {\cal J}^{\mu\nu}_{R\,\lambda}}{\partial p_\tau}\varphi^{(L)\rho}_{(L)\tau}=\frac{1}{\Lambda}\left(n^\mu  {\cal J}^{\nu\rho}_{R\,\lambda}-n^\nu  {\cal J}^{\mu\rho}_{R\,\lambda}\right)\,.
    \label{diffeqjr}
\end{equation}
A similar result can be obtained from the Poisson bracket between the Lorentz generator and the right-particle space-time coordinates, viz.
\begin{equation}
	\frac{\partial {\cal J}^{\mu\nu}_{L\,\lambda}}{\partial q_\tau}\varphi^{(R)\rho}_{(R)\tau}=\frac{1}{\Lambda}\left(n^\mu  {\cal J}^{\nu\rho}_{L\,\lambda}-n^\nu  {\cal J}^{\mu\rho}_{L\,\lambda}\right)\,.
     \label{diffeqjl}
\end{equation}
Both equations clearly show that the Lorentz transformations of the left (right) particle must depend on the right (left) momenta. This implies that these transformations cannot be obtained from the usual linear Lorentz transformations through a change of momentum basis. This differs from the starting idea of DSR models \cite{Amelino-Camelia2001,Amelino-Camelia2002a}, where the composition law of momenta can be obtained through a change in momentum basis, and the coproduct of Lorentz generators does not mix momenta. 

For the functions $\varphi$ in Eq.~\eqref{eq:varphi}, we obtain a complex system of differential equations, given by Eqs. \eqref{eq_composition_Lorentz2}, \eqref{diffeqjl}, and \eqref{diffeqjr}. Then, using the Lorentz transformations in Eq.~\eqref{eq:jlight}, we can solve it order-by-order, finding (up to the second order): 
\begin{equation}
	\begin{split}
  {\cal J}^{\mu\nu}_{L\,\lambda}(p,q)=  &  \left( \delta^{\mu}_{\lambda} p^{\nu} - \delta^{\nu}_{\lambda} p^{\mu} \right) 
    - \left(\frac{1}{\Lambda} \left(\frac{1}{2} p^2-pq\right)+\frac{1}{\Lambda^2}\left(\frac{1}{2}  p^2(pn - qn)+pq(pn+qn)\right)\right) \left( \delta^{\mu}_{\lambda} n^{\nu} - \delta^{\nu}_{\lambda} n^{\mu} \right) \\
    & + \frac{pq}{\Lambda^2} n_{\lambda} \left( n^{\mu} p^{\nu} - n^{\nu} p^{\mu} \right) - \frac{p_{\lambda} qn}{\Lambda^2} \left( n^{\mu} p^{\nu} - n^{\nu} p^{\mu} \right)   + \frac{ q_{\lambda}}{\Lambda}\left(-1+\frac{qn}{\Lambda}\right) \left( n^{\mu} p^{\nu} - n^{\nu} p^{\mu} \right)+\mathcal{O}\left(\frac{1}{\Lambda^3}\right) .\\
     {\cal J}^{\mu\nu}_{R\,\lambda}(p,q)=& {\cal J}^{\mu\nu}_{L\,\lambda}(q,p)\,.
\end{split}
\end{equation}
It is important to note that
\begin{equation}
	  {\cal J}^{\mu\nu}_{L\,\lambda}(p,0)=   {\cal J}^{\mu\nu}_{\lambda}(p)\,,\qquad  {\cal J}^{\mu\nu}_{R\,\lambda}(0,q)=   {\cal J}^{\mu\nu}_{\lambda}(q)\,,
\end{equation}
and that Eq.~\eqref{eq:cplusjs} is satisfied.

\section{Second attempt of implementation of locality}
\label{sec:second}
We now consider the second implementation of the locality of interactions given by Eq.~\eqref{loc2} as follows. In this case, as shown in~\cite{Carmona:2019vsh}, the noncommutative functions are given by
\begin{equation}
	\varphi^{(L)\alpha}_{(L)\nu}(p,q)=	\varphi^{\alpha}_{\nu}(p)\,,\qquad \varphi^{(L)\alpha}_{(R)\nu}(p,q)=	\varphi^{\alpha}_{\nu}(q)-L^\alpha_{\nu}(q)\,,\qquad 	\varphi^{(R)\alpha}_{(L)\nu}(p,q)=	\varphi^{\alpha}_{\nu}(p)-R^\alpha_{\nu}(p)\,,\qquad \varphi^{(R)\alpha}_{(R)\nu}(p,q)=	\varphi^{\alpha}_{\nu}(q)\,.
    \label{loc22}
\end{equation}
With this implementation of locality, the composition law must be associative~\cite{Carmona:2019vsh}. Let us examine this case in greater detail.

\subsection{Composition law}
As in the first attempt, we study the composition law from the Poisson bracket of both noncommutative coordinates:
\be
\begin{split}
\lbrace\tilde{y}_L^\mu, \tilde{z}_R^\nu\rbrace = &y^\lambda \,\left(\frac{\partial\varphi^\mu_\lambda(p)}{\partial p_\sigma}\left(\varphi^\nu_\sigma(p)-R^\nu_\sigma(p)\right)-\left(\frac{\partial\varphi^\nu_\lambda(p)}{\partial p_\sigma}-\frac{\partial R^\nu_\lambda(p)}{\partial p_\sigma}\right)\varphi^\mu_\sigma(p)\right)\\
+ & z^\lambda \,\left(\left(\frac{\partial\varphi^\mu_\lambda(q)}{\partial p_\sigma}-\frac{\partial L^\mu_\lambda(q)}{\partial p_\sigma}\right)\varphi^\nu_\sigma(q)-\frac{\partial\varphi^\nu_\lambda(q)}{\partial p_\sigma} \left(\varphi^\mu_\sigma(q)-L^\mu_\sigma(q)\right)\right)\\
=&y^\lambda\left(C^{\mu\nu}_\rho \varphi^\rho_\lambda(p)-\frac{\alpha}{\Lambda^2}{\cal J}^{\mu\nu}_{L\,\lambda} +\varphi^\mu_\sigma \frac{\partial R^\nu_\lambda(p)}{\partial p_\sigma}-\frac{\partial\varphi^\mu_\lambda(p)}{\partial p_\sigma}R^\nu_\sigma(p)\right)\\
+&z^\lambda\left(C^{\mu\nu}_\rho \varphi^\rho_\lambda(q)-\frac{\alpha}{\Lambda^2}{\cal J}^{\mu\nu}_{R\,\lambda}-\frac{\partial L^\mu_\lambda(q)}{\partial p_\sigma}\varphi^\nu_\sigma(q)+\frac{\partial\varphi^\nu_\lambda(q)}{\partial p_\sigma} L^\mu_\sigma(q)\right) \,.
\label{secondyz}
\end{split}
\ee
In the last step, we used the Poisson bracket of the one-particle system of the first equation of~\eqref{1symmetric_algebra_alpha}. 
Now, since we want~\eqref{symmetric_algebra_alpha} to hold, Eq.~\eqref{secondyz} must also be equal to
\begin{equation}
	\lbrace\tilde{y}_L^\mu, \tilde{z}_R^\nu\rbrace = y^\lambda \left(C^{\mu\nu}_{L\,\rho} \varphi^\rho_\lambda(p)-C^{\mu\nu}_{R\,\rho} \left(\varphi^\rho_\lambda(p)-R^\rho_\lambda(p)\right)-\frac{\alpha}{\Lambda^2}{\cal J}^{\mu\nu}_{L\,\lambda}\right)+z^\lambda \left(-C^{\mu\nu}_{R\,\rho} \varphi^\rho_\lambda(p)+C^{\mu\nu}_{L\,\rho} \left(\varphi^\rho_\lambda(p)-L^\rho_\lambda(p)\right)-\frac{\alpha}{\Lambda^2}{\cal J}^{\mu\nu}_{L\,\lambda}\right)\,.
    \label{secondyz2}
\end{equation}
By equating Eqs.~\eqref{secondyz} and~\eqref{secondyz2}, and using Eq.~\eqref{eq:condition_Cs}, we find 
\begin{equation}
\begin{split}
    \frac{\partial R^{\nu}_{\lambda}(p)}{\partial p_\sigma}\varphi^\mu_\sigma(p)-\frac{\partial \varphi^\mu_\lambda(p)}{\partial p_\sigma}R^{\nu}_{\sigma}(p)=&C^{\mu\nu}_{R\,\rho} R^{\rho}_{\lambda}(p)\,,\implies\\
     \frac{\partial R^{\mu}_{\lambda}(p)}{\partial p_\sigma}\varphi^\nu_\sigma(p)-\frac{\partial \varphi^\nu_\lambda(p)}{\partial p_\sigma}R^{\mu}_{\sigma}(p)=&C^{\mu\nu}_{L\,\rho} R^{\rho}_{\lambda}(p)\,,\\
       \frac{\partial L^{\mu}_{\lambda}(q)}{\partial q_\sigma}\varphi^\nu_\sigma(q)-\frac{\partial \varphi^\nu_\lambda(q)}{\partial q_\sigma}L^{\mu}_{\sigma}(q)=&C^{\mu\nu}_{L\,\rho} L^{\rho}_{\lambda}(q)\,,
\end{split}
\end{equation}
where in the second line, we used Eq.~\eqref{eq:clcr}. Then, we see that the same equation for $L$ and $R$ must be satisfied, and thus, they are the same functions. Hence, the composition law must be symmetrical. As a particular example, we can consider the sum. In this case, the previous equation becomes
\begin{equation}
\begin{split}
         -\frac{\partial \varphi^\nu_\lambda(p)}{\partial p_\mu}=&C^{\mu\nu}_{L\,\lambda}=-\delta^\nu_\lambda n^\mu+\eta^{\mu\nu}n_\lambda\,,
\end{split}
\end{equation}
so 
\begin{equation}
	\varphi^\mu_\nu(p)=\delta^\mu_\nu \left(1+\frac{pn}{\Lambda}\right)-\frac{n_\nu p^\mu}{\Lambda}\,.
\end{equation}

\subsection{Lorentz transformations}
From the Poisson bracket of the Lorentz generators and the noncommutative space-time coordinates, Eq.~\eqref{eq:jvarphi}, we have
\begin{equation}
		{\cal J}^{\mu\nu}_{\lambda}=\delta^\mu_\lambda k^\nu-\delta^\nu_\lambda k^\mu\,.
\end{equation}
Moreover, when using the Poisson brackets involving the Lorentz generators, one finds the same ones as SR, that is,
\begin{equation}
	{\cal J}^{\mu\nu}_{L\,\lambda}=\delta^\mu_\lambda p^\nu-\delta^\nu_\lambda p^\mu\,,\qquad {\cal J}^{\mu\nu}_{R\,\lambda}=\delta^\mu_\lambda q^\nu-\delta^\nu_\lambda q^\mu\,.
\end{equation}
This means that one can choose an implementation of locality of interactions with the same kinematics as SR, but with a noncommutative spacetime. This differs from the first attempt, where the Lorentz transformations in the two-particle system involve both momenta. However, both implementations led to an associative and symmetric composition law of momenta.

\section{General implementation of locality}
\label{sec:general}
In this case, it is not possible to obtain analytical expressions and relationships between the composition law, the Lorentz generators, and the non-commutative coordinates. However, for the lightlike case, we show that the implementation of locality by preserving the Poisson brackets of~\eqref{symmetric_algebra_alpha} is feasible order-by-order. Indeed, we explicitly write the results up to the first order in the power-series expansion in $\Lambda$. 

By considering the same $\varphi$ function of Eq.~\eqref{eq:varphi}, 
the Lorentz generator in the one-particle system is given by Eq.~\eqref{eq:jlight}. Now, using the conditions~\eqref{eq_composition_Lorentz2},~\eqref{eq:varphi_L2},~\eqref{loc3},  and~\eqref{symmetric_algebra_alpha}, with a generic composition law at fist order,
\begin{equation}
	(p \oplus q)_\mu=p_{\mu} \left( 1 + \beta_1 \frac{q n}{\Lambda} \right) 
    + q_{\mu} \left( 1 + \beta_2 \frac{p n}{\Lambda} \right) 
    + n_{\mu} \left( \beta_3 \frac{ p q}{\Lambda} +\beta_4  \frac{p n \, q n}{\Lambda} \right)\,,
\end{equation}
the Lorentz transformations are given by 
\begin{equation}
\begin{aligned}
  {\cal J}^{\mu\nu}_{L\,\lambda}(p,q)=  & \left( \delta^{\mu}_{\lambda} p^{\nu} - \delta^{\nu}_{\lambda} p^{\mu} \right) 
   + (\tau_1-\beta_4) n_{\lambda} \left(\frac{qn}{\Lambda}  \left( n^{\mu} p^{\nu} - n^{\nu} p^{\mu} \right) 
     + \frac{pn}{\Lambda}  \left( n^{\mu} q^{\nu} - n^{\nu} q^{\mu} \right) \right)\\
       - &  (\tau_2+\beta_1)\frac{p_{\lambda}}{\Lambda} \left( n^{\mu} q^{\nu} - n^{\nu} q^{\mu} \right)- 
 (\tau_3+\beta_2)\frac{q_{\lambda}}{\Lambda} \left( n^{\mu} p^{\nu} - n^{\nu} p^{\mu} \right)\,,\\
 {\cal J}^{\mu\nu}_{R\,\lambda}(p,q)=  & \left( \delta^{\mu}_{\lambda} q^{\nu} - \delta^{\nu}_{\lambda} q^{\mu} \right) 
   - \tau_1 n_{\lambda} \left(\frac{qn}{\Lambda}  \left( n^{\mu} p^{\nu} - n^{\nu} p^{\mu} \right) 
     + \frac{pn}{\Lambda}  \left( n^{\mu} q^{\nu} - n^{\nu} q^{\mu} \right) \right)\\
       + & \tau_2\frac{p_{\lambda}}{\Lambda} \left( n^{\mu} q^{\nu} - n^{\nu} q^{\mu} \right)- 
 \tau_3\frac{q_{\lambda}}{\Lambda} \left( n^{\mu} p^{\nu} - n^{\nu} p^{\mu} \right) \,.
\end{aligned}
\end{equation}
Finally, the $\varphi$ functions are given by
\begin{equation}
\begin{aligned}
  \varphi^{(L)\mu}_{(L)\nu}(p,q)   &= \delta^{\mu}_\nu \left(1+\frac{pn}{\Lambda}-(\tau_2+\beta_1)\frac{qn}{\Lambda}\right)
   + (\tau_1-\beta_4)\frac{qn}{\Lambda}
     n^{\mu} n_{\nu}  
    -(\tau_3+\beta_2) \frac{n^{\mu} q_{\nu}}{\Lambda} 
    + (1-\beta_3-\tau_4)\frac{q^{\mu} n_{\nu}}{\Lambda} \,,\\
      \varphi^{(L)\mu}_{(R)\nu}(p,q)   &= \delta^{\mu}_\nu (1+\tau_2)\frac{qn}{\Lambda} 
    -\tau_1 \frac{qn}{\Lambda} n^{\mu} n_{\nu} +\tau_3 \frac{n^{\mu} q_{\nu}}{\Lambda}  
   +(\tau_4-1)\frac{q^{\mu} n_{\nu}}{\Lambda}\,, \\
      \varphi^{(R)\mu}_{(L)\nu}(p,q)   &= \delta^{\mu}_\nu (1-\tau_3-\beta_2)\frac{pn}{\Lambda}
    -(\tau_2+\beta_1) \frac{n^{\mu} p_{\nu}}{\Lambda} 
    -(\beta_3+\tau_4) \frac{p^{\mu} n_{\nu}}{\Lambda} 
    +(\tau_1-\beta_4)\frac{pn}{\Lambda} n^{\mu} n_{\nu}\,, \\
      \varphi^{(R)\mu}_{(R)\nu}(p,q)   &= \delta^{\mu}_\nu \left(1+\tau_3\frac{pn}{\Lambda}+\frac{qn}{\Lambda}\right)
    + \tau_2 \frac{n^{\mu} p_{\nu}}{\Lambda}  
    + \tau_4 \frac{p^{\mu} n_{\nu}}{\Lambda}  
    -\tau_1 \frac{pn}{\Lambda} n^{\mu} n_{\nu} \,.
\end{aligned}
\end{equation}
It can be observed that in addition to the four independent coefficients of the composition law ($\beta$s), four more appear in the Lorentz transformations and in the $\varphi$ functions ($\tau$s). 

A possible restriction that can be implemented in this generic approach is to impose the kinematics of $\kappa$-Poincaré, for which the composition law must be symmetric and the Lorentz transformation of the left particle should not depend on the right momentum. This will correspond to
\begin{equation}
\beta_1=\beta_3=\beta_4=\tau_1=\tau_2=0\,,\qquad \beta_2=-\tau_3=1\,.
\end{equation}
Even with this restriction, the $\varphi$ functions are not completely determined, because $\tau_4$ is completely free. 

It seems that, with this generic implementation of locality, we can make any composition law compatible with a noncommutativity of space-time coordinates forming a Poisson-Lie algebra together with the Lorentz generators. Then, this can accommodate different kinematics, such as $\kappa$-Poincaré, Snyder, and hybrid models, as well as those less studied in the literature~\cite{Ballesteros_1994,Ballesteros:2003kz}. We hope to explore in future works how to impose more mathematical or physical restrictions to univocally determine the $\varphi$ functions, the composition law, and the Lorentz generators.

\section{New geometrical interpretation}
\label{sec:geometry}
Now, we explore a different geometric interpretation of that proposed in~\cite{Relancio:2021ahm} of the locality of interactions. We introduce the notation
\begin{equation}
	\left(\tilde y^\rho-\tilde z^\rho\right)\eta_{\rho\sigma}\left(\tilde y^\sigma-\tilde z^\sigma\right)= X^A G_{AB}(P)X^B\,,
\end{equation}
where $A,\,B$ run from 0 to 7, $X^A=(y^\mu,z^\mu)$,  $P_A=(p_\mu,q_\mu)$, and
\begin{equation}
G_{AB}(P)=
\begin{pmatrix}
g^{LL}_{\mu\nu}(p,q) & g^{LR}_{\mu\nu}(p,q) \\
g^{RL}_{\mu\nu}(p,q) & g^{RR}_{\mu\nu}(p,q) 
\end{pmatrix}\,,
\label{eq:8-metric2}
\end{equation}  
with 
\begin{equation}
\begin{split}
g^{LL}_{\mu\nu}(p,q)&=\left(\varphi^{(L)\alpha}_{(L)\mu}(p,q)-\varphi^{(R)\alpha}_{(L)\mu}(p,q)\right)\eta_{\alpha\beta}\left(\varphi^{(L)\beta}_{(L)\nu}(p,q)-\varphi^{(R)\beta}_{(L)\nu}(p,q)\right)\,,\\
g^{LR}_{\mu\nu}(p,q)&=g^{RL}_{\nu\mu}(p,q)=
\left(\varphi^{(L)\alpha}_{(L)\mu}(p,q)-\varphi^{(R)\alpha}_{(L)\mu}(p,q)\right)\eta_{\alpha\beta}\left(\varphi^{(L)\beta}_{(R)\nu}(p,q)-\varphi^{(R)\beta}_{(R)\nu}(p,q)\right)\,,\\
g^{RR}_{\mu\nu}(p,q)&=\left(\varphi^{(L)\alpha}_{(R)\mu}(p,q)-\varphi^{(R)\alpha}_{(R)\mu}(p,q)\right)\eta_{\alpha\beta}\left(\varphi^{(L)\beta}_{(R)\nu}(p,q)-\varphi^{(R)\beta}_{(R)\nu}(p,q)\right)\,.
\end{split}
\label{eq:metric_tetrad2}
\end{equation}

We can now study some properties of this metric.
Starting from the Poisson brackets~\eqref{symmetric_algebra_alpha}, one can note that
\begin{equation}
	\lbrace X^A G_{AB}(P)X^B,J^{\mu\nu}\rbrace=0\,.
\end{equation}
This can also be obtained from Eq.~\eqref{eq:rj}. Moreover, from Eq.~\eqref{eq:rcomp} one finds
\begin{equation}
	\lbrace  X^A G_{AB}(P)X^B, \left(p\oplus q\right)_\mu \rbrace=0\,.
\end{equation}
Then, automatically, the composition law and Lorentz transformations in the two-particle system become isometries of the metric~\eqref{eq:metric_tetrad2}.

This means that one can construct a multi-particle metric in phase space that is different from that in Eq.~\eqref{eq:metric_tetrad}. Interestingly, this metric is degenerate (non-invertible) in the zero-momentum limit. This can be easily seen just by replacing  
\begin{equation}
	\varphi^{(L)\alpha}_{(L)\mu}(p,q)=\varphi^{(R)\alpha}_{(R)\mu}(p,q)=\delta^\alpha_\mu\,,\qquad \varphi^{(L)\alpha}_{(R)\mu}(p,q)=\varphi^{(L)\alpha}_{(R)\mu}(p,q)=0\,,
\end{equation}
in Eq. \eqref{eq:metric_tetrad2}, finding that the metric is
\begin{equation}
G=\begin{pmatrix}
    \eta & -\eta \\
    -\eta & \eta
\end{pmatrix}\,.
\end{equation}
Thus, a power series expansion on the high-energy scale $\Lambda$ is not possible. This metric could be not degenerate for the complete expressions of $\varphi$ or for $\lambda_1\neq \lambda_2$. A different option is to consider a non-symmetric metric. The study of this geometrical realisation is beyond the scope of this paper and will be studied elsewhere.

\section{Conclusions}
\label{sec:conclusions}
In this work, a description of the different Poisson-Lie algebras formed by noncommutative space-time coordinates and Lorentz generators is presented. In particular, we considered an algebra formed by these 14 generators together with a fixed vector, which can be timelike, lightlike, or spacelike. These different possibilities lead  to three different families depending on the square of the fixed vector, all of which correspond to the $R^{3,1}\rtimes R^{3,1}\rtimes O(3,1)$ algebra. In this paper we focused on the particular case in which there is a symmetry between both particles, thus having a smooth limit when only one particle is considered. Moreover, we showed how to extend this algebra to any number of particles. The study of the nonsymmetric case is left for future work.

As discussed in~\cite{Lizzi:2021rlb}, the study of the algebra of a multi-particle system can be applied to a QFT with a deformed relativistic kinematics. With this work, we go beyond the proposal of~\cite{Lizzi:2021rlb} because we also study the timelike and spacelike cases. These deformations could also be applied to a QFT and any number of particles.

Next, we consider how to implement a locality of interactions compatible with such a Poisson-Lie algebraic structure. We studied three different implementations of locality, including some particular cases and the most general one. For the former, we find that the composition law of the momenta must be symmetric and associative, so it can be obtained from the sum through a change of momentum basis. For the latter case, analytical expressions cannot be obtained, so a power series expansion for small momenta is performed. In this general case, any desired kinematics can be constructed, including $\kappa$-Poincaré. 

Finally, we explore a geometrical interpretation of the implementation of locality compatible with the 14-dimensional Poisson-Lie algebra. As in~\cite{Relancio:2021ahm}, one can consider the functions leading to a noncommutative spacetime as the tetrad of an eight-dimensional metric in the phase space of two particles. This metric is degenerate in the zero-momentum limit; therefore, a power series expansion cannot be considered when studying its properties. 
It is interesting to note that degenerate metrics appear in the loop approach to canonical quantum Einstein gravity~\cite{Jacobson:1987qk} as well as in the context of Kähler~\cite{Tosatti_2009} and Finsler~\cite{NGUYEN2021112247} metrics. In future work, we hope to explore the features of this geometrical setup and compare it with previous studies in the literature, such as the connection with the noncommutativity of geodesics studied in~\cite{Ballesteros:2022nyx}.

\paragraph{Acknowledgments.}
The author thanks José Manuel Carmona and José Luis Cortés for their collaboration during the first stages of this work, and also Ángel Ballesteros and Lucía Santamaría for useful comments. This work has been supported by the grant PID2023-148373NB-I00 funded by MCIN /AEI /10.13039/501100011033 / FEDER, UE, and by the Q-CAYLE Project funded by the Regional Government of Castilla y Le\'on (Junta de Castilla y Le\'on) and by the Ministry of Science and Innovation (MCIN) through the European Union funds NextGenerationEU (PRTR C17.I1). The author would like to acknowledge the contribution of the COST Action CA23130. The author has benefited from the activities of COST Action CA23115:
Relativistic Quantum Information, funded  by COST (European Cooperation in Science and Technology).

\end{document}